\documentclass[11pt]{article}
\pdfoutput=1
\usepackage{a4wide}
\usepackage{amsmath,amssymb,amsfonts}
\usepackage{comment}
\usepackage{epsfig,verbatim,graphics}
\usepackage{subfigure,slashed,siunitx}
\usepackage[table,x11names]{xcolor}

\usepackage{hyperref}

\newcommand{\mathsym}[1]{{}}

\newcommand{\eref}[1]{(\ref{#1})}

\renewcommand\({\left(}
\renewcommand\){\right)}
\renewcommand\[{\left[}
\renewcommand\]{\right]}

\newcommand{\dd}{{\rm d}}
\declareslashed{}{{^-}}{.1}{0.1}{\rm d}

\newcommand{\e}{{\rm e}}

\newcommand\eps{\epsilon}
\newcommand\mpl{m_{\rm P}}

\def\ba{\begin{eqnarray}}
\def\ea{\end{eqnarray}}
\def\be{\begin{equation}}
\def\ee{\end{equation}}

\def\L{\mathcal{L}}
\def\O{\mathcal{O}}

\def\nn{\nonumber}
\def\({\left(}
\def\){\right)}

\def\eref#1{(\ref{#1})}

\usepackage{tikz}
\usetikzlibrary{arrows,shapes}
\usetikzlibrary{trees}
\usetikzlibrary{matrix,arrows} 				% For commutative diagram
\usetikzlibrary{positioning}				% For "above of=" commands
\usetikzlibrary{calc,through}				% For coordinates
\usetikzlibrary{decorations.pathreplacing}  % For curly braces
\usepackage{pgffor}							% For repeating patterns

\usetikzlibrary{decorations.pathmorphing}	% For Feynman Diagrams
\usetikzlibrary{decorations.markings}
\tikzset{
    vector/.style={decorate, decoration={snake}, draw},
	provector/.style={decorate, decoration={snake,amplitude=2.5pt}, draw},
	antivector/.style={decorate, decoration={snake,amplitude=-2.5pt}, draw},
    fermion/.style={draw=black, postaction={decorate},
        decoration={markings,mark=at position .55 with {\arrow[draw=black]{>}}}},
    fermionbar/.style={draw=black, postaction={decorate},
        decoration={markings,mark=at position .55 with {\arrow[draw=black]{<}}}},
    fermionnoarrow/.style={draw=black},
    gluon/.style={decorate, draw=black,
        decoration={coil,amplitude=4pt, segment length=5pt}},
    scalar/.style={dashed,draw=black, postaction={decorate},
        decoration={markings,mark=at position .55 with {\arrow[draw=black]{>}}}},
    scalarbar/.style={dashed,draw=black, postaction={decorate},
        decoration={markings,mark=at position .55 with {\arrow[draw=black]{<}}}},
    scalarnoarrow/.style={dashed,draw=black},
    electron/.style={draw=black, postaction={decorate},
        decoration={markings,mark=at position .55 with {\arrow[draw=black]{>}}}},
	bigvector/.style={decorate, decoration={snake,amplitude=4pt}, draw},
}

\newcommand{\roughly}[1]{\mathrel{\raise.3ex\hbox{$#1$\kern-0.85em
\lower1ex\hbox{$\sim$}}}}

\begin{document}
\begin{titlepage}%1
\begin{center}

\today
\hfill Nikhef 2016-009

\vskip 1.5cm

{\LARGE \bf UV (in)sensitivity of Higgs inflation}

\vskip 1cm

% this command makes the email footnote's into symbols
\renewcommand\star{\thefootnote}{\fnsymbol{footnote}}
\setcounter{footnote}{0}

{\bf
Jacopo Fumagalli\footnote{{\tt jacopof@nikhef.nl },
        \footnotemark[2]{\tt mpostma@nikhef.nl}} and
    Marieke Postma\footnotemark[2]
}

% this command reverts the footnote style to back to number
\renewcommand*{\thefootnote}{\number{footnote}}
\setcounter{footnote}{0}

\vskip 25pt

{\em 
Nikhef, \\Science Park 105, \\1098 XG Amsterdam, The Netherlands
}

\end{center}

\vskip 0.5cm

\begin{center} {\bf ABSTRACT}\\[3ex]\end{center} 
%In (Standard Model)
%Higgs inflation, unknown UV physics modifies the running of the
%couplings, and thereby changes the shape of the inflationary potential.
%We show that, nevertheless, the inflationary predictions are to a very
%good approximation unaffected by the renormalization group
%evolution. This implies in general that one cannot predict the spectral index and
%tensor-to-scalar ratio from precise top and Higgs mass measurements at
%the LHC, nor can one probe effects of UV physics on the running.
The predictions of Standard Model Higgs inflation are in excellent
agreement with the Planck data, without the need for new fields.
However, consistency of the theory requires the presence of (unknown)
threshold corrections. These modify the running of the couplings, and
thereby change the shape of the inflationary potential. This raises
the question how sensitive the CMB parameters are to the UV
completion. We show that, due to a precise cancellation, the
inflationary predictions are almost unaffected.  This implies in
general that one cannot relate the spectral index and tensor-to-scalar
ratio to the precise top and Higgs mass measurements at the LHC, nor
can one probe effects of UV physics on the running.

\end{titlepage}

\newpage
\setcounter{page}{1} \tableofcontents

\newpage

%%%%%%%%%%%%%%%%%%%%%%%%%%%%%%%%%%%%%%%%%%%%%%%%%%%%%%
%%%%%%%%%%%%%%%%%%%%%%%%%%%%%%%%%%%%%%%%%%%%%%%%%%%%%%
%%%%%%%%%%%%%%%%%%%%%%%%%%%%%%%%%%%%%%%%%%%%%%%%%%%%%%

\section{Introduction}

Standard Model Higgs inflation \cite{fakir,salopek,
  bezrukov1,Bezrukov:2013fka} has attracted much attention over the
last years. This is not surprising, as the model has many appealing
features --- at least, at the classical level.  With the Higgs field
as inflaton the model is firmly embedded in the Standard Model (SM).
Only one new interaction is needed, a non-minimal coupling of the
Higgs field to the Ricci tensor, making the set-up minimal.  And
finally, the predictions for the inflationary observables are in
excellent agreement with the latest Planck data \cite{Planck}

Although this minimal approach is attractive, it is not clear whether
Higgs inflation is fully consistent as a quantum theory.  First,
including the running of couplings, the potential may become unstable
at energy scales below the inflationary scale. For the best fit values
of the top and Higgs mass this indeed happens, but it should be noted
that vacuum stability of the SM up to the Planck scale is only
excluded at the 2-3$\sigma$ level \cite{disc1, disc2,
  branchina,branchina2, archil, alexss,kniehl}.  Furthermore, with
extra matter, e.g. a dark matter particle, the instability bound can
be evaded \cite{lebedev,espinosa,Basso:2013nza}.

A second issue with quantum Higgs inflation is the unitarity bound
\cite{burgess1,barbon,burgess2,Hertzberg1,bezrukov4,linde_higgs,
  cliffnew,He2}. Tree level unitarity is lost at energies well below
the Planck scale, and new degrees of freedom  \cite{giudice,mirage} or strong dynamics
\cite{ufuk,calmet}) should become important at this scale.
Although the energy scale of the inflationary potential is always
below the field-dependent unitarity cutoff
\cite{bezrukov4,linde_higgs,moss,He1}, which makes the semiclassical
approximation meaningful, this is not so for the field value.  To
forbid non-renormalizable operators that spoil the inflationary
potential already at the classical level, an (approximate) shift
symmetry has to be assumed. This is no different from chaotic
inflation.
%\textcolor{red}{Moreover, for the only self consistency of
%  HI, we do not need to introduce such operators as we explain in the
%  following.}

Thirdly, the theory is not renormalizable over the full field
range. It has been shown that for small, mid and large field values
Higgs inflation is renormalizable in the usual effective field theory
(EFT) sense \cite{damien2} (see also earlier work \cite{bezrukov_loop,
  bezrukov3, wilczek, barvinsky, barvinsky2,
  barvinsky3,kyle,damien,GMP}). However, these EFTs need to be patched
together at the boundaries of the different field regimes, and it is
here that we expect non-renormalizable operators to become important.
We will also refer to these higher order operators as threshold
corrections, and more generically, speak about threshold corrections
to the renormalization group equations (RGEs) or to the inflationary
observables, meaning the effect of the higher order operators on these
quantities.

Thus for a consistent quantum field theory, new physics is needed
below the Planck scale. This begs the question: how sensitive is Higgs
inflation to the unknown UV physics~\cite{cliffnew,Hertzberg2}?  If
the model predictions demanded a particular UV completion it would
mean that the simplicity of the set-up, to which it owes much of its
success, would be completely spoiled.  In this paper we will show that
as long as the UV corrections do not affect the inflaton potential at
tree level, but only enter at loop level via corrections to the
renormalization group equations, the inflationary predictions are to a
very good approximation unaffected. Indeed, whatever the exact running
of the couplings, the spectral index $n_s$ and tensor-to-scalar ratio
$r$ have at leading order in the slow roll expansion a universal
value:~\footnote{This is for inflation on the flat plateau of the
  potential, as is usually meant by ``Higgs inflation'' (and at tree
  level is the only possibility).  For some fine-tuned values of the
  couplings, inflation near a maximum or inflection point of the
  potential is possible; in this latter case, the predictions are
  sensitive to the details of the potential, and thus to the unknown
  UV physics. }
\be
n_s = 1 - \frac{2}{N_\star} + \O(N_\star^{-2})\simeq 0.967 ,\qquad
r = \frac{12}{N_\star^2} + \O(N_\star^{-3}, \xi^{-1})\simeq 0.003
\label{universal}
\ee
with $N_\star \approx 60$ the number of efolds of observable
inflation, and $\xi \gg 1$ the non-minimal coupling. We can rephrase
our statement as follows: as long as the non-minimally coupled Higgs
is a viable inflaton candidate --- no large tree-level UV corrections
to the potential, and no RGE induced instability of the potential ---
the predictions are extremely robust and to high precision are equal
to the tree level results, in excellent agreement with the Planck data
on the CMB \cite{Planck}.

This paper is organized as follows.  In the next section we first
describe the full quantum action for Higgs inflation. To set the
notation, in section \ref{s:L_tree} we introduce the classical action.
In section \ref{s:UV} we discuss the UV completion of the theory.
Following \cite{cliffnew} we introduce a set of threshold corrections,
which we will use in our numerical results.  The particular set of
non-renormalizable operators can be motivated in two ways, either by
assuming an approximate shift symmetry, or by demanding that UV
physics only enters where needed for the consistency of the theory,
namely at the boundary of the small and mid field regimes. We stress,
however, that the universal results for $n_s$ and $r$ as given in
\eref{universal} do not depend on this choice. There has been some
debate in the literature on frame dependence and the choice of
renormalization scale \cite{bezrukov_loop, bezrukov3, wilczek,
  barvinsky, barvinsky2, barvinsky3,kyle}. In section \ref{s:mu} we
will argue that this choice is unambiguously defined
\cite{damien,volpe}. To end this section, in \ref{s:RGE} we shortly
discuss the renormalization group equations (RGEs) and
also give details on the numerical implementation.

In section \ref{s:inflation} we then turn to the predictions for
inflation. In \ref{s:flat} we calculate the inflationary observables
for Higgs inflation, using the RGE improved potential which includes
the effects of running couplings.  We show analytically that the
spectral index and scalar-to-tensor ratio are to first order in an
$1/N_\star$ expansion insensitive to the running of the couplings. To
investigate the possibility of hilltop inflation, inflation near a
maximum of the potential, we turn to our numerical results. As
discussed in section \ref{s:max}, we find that for fine-tuned boundary
conditions (the top/Higgs mass values at the electroweak scale, and
the Wilson coefficients of the non-renormalizable operators) hilltop
inflation is possible. Since the potential near the maximum is tuned
to be flat enough for 60 efolds of inflation, it comes as no surprise
that this tuning depends very sensitively on the details of the
potential, and thus also on the running of the couplings.

We end in \ref{s:conclusions} with some concluding remarks.

Our sign convention for the metric is mostly positive $(-,+,+,+)$.
The dependence on the Planck mass is kept explicitly only in the first
part where we introduce the model and discuss the unitary cutoff, in
the rest of the paper we set the reduced Planck mass to unity
$\mpl=(\sqrt{8\pi G})^{-1}=1$.

%%%%%%%%%%%%%%%%%%%%%%%%%%%%%%%%%%%%%%%%%%%%%%%%%%%%%%
%%%%%%%%%%%%%%%%%%%%%%%%%%%%%%%%%%%%%%%%%%%%%%%%%%%%%%
%%%%%%%%%%%%%%%%%%%%%%%%%%%%%%%%%%%%%%%%%%%%%%%%%%%%%%

\section{Effective action for Higgs inflation}
\label{s:action}

In this section we discuss the effective action for Higgs inflation;
in section \ref{s:inflation} we then calculate the inflationary
observables for this action. 

The loop corrections can be incorporated in an RGE improved action
with running couplings.  We include a class of threshold corrections
(coming from the UV completion) which only enter the inflationary
physics via their effect on the renormalization group equations
(RGEs). Finally, we discuss the choice of renormalization scale.

%%%%%%%%%%%%%%%%%%%%%%%%%%%%%%%%%%%%%%%%%%%%%%%%%%%%%%

\subsection{Higgs inflation in Einstein frame}
\label{s:L_tree}

To set the notation, let's start with defining the classical action for Higgs inflation in
the Jordan frame: 
\be
\L_J = \sqrt{-g^J}\[ \frac12 \mpl^2 \(1+ \frac{2\xi \Phi^\dagger
  \Phi}{\mpl^2} \) R[g^J] + \L_{\rm SM}\] .
\label{L_jordan}
\ee
with $\Phi$ the standard model Higgs doublet and $\xi$ the non-minimal
coupling to gravity.  The gravitational action can be brought in
Einstein-Hilbert form by a conformal transformation
$g_{\mu\nu} = \Omega^2 g_{\mu \nu}^J$ with conformal factor
\be
\Omega^2 = \(1+ \frac{2\xi \Phi^\dagger \Phi}{\mpl^2}\) .
\label{Omega}
\ee 
The resulting Einstein frame action is
\be
\L_E =\sqrt{-g}\[ \frac12 \mpl^2 R[g]
-  \frac{1}{\Omega^2} (\partial_\mu \Phi)^\dagger (\partial^\mu \Phi) - \frac{3 \xi^2}{\mpl^2\Omega^4}
 \partial_\mu (\Phi^\dagger \Phi) \partial^\mu (\Phi^\dagger \Phi) - \frac{V_J}{\Omega^4}+...\] ,
\label{LE0}
\ee
with $ V_J = \lambda(\Phi^\dagger \Phi - v^2/2)^2 $. The Lagrangian
for the classical background field $\Phi= \frac1{\sqrt{2}}
\( \begin{array}{c} 0 \\ \phi \end{array} \) $ is
\be
\L_E =\sqrt{-g}\[ \frac12 \mpl^2 R[g]
 -\frac12 \gamma_{\phi\phi}(\phi) (\partial \phi)^2 - \frac{\lambda
  ( \phi^2 -v^2)^2}{4\Omega^4} \] .
\label{LE}
\ee
In the large field limit $\phi^2 \gg \mpl^{2}/\xi$ the potential approaches a
constant value developing a flat plateau that can support a
period of slow roll inflation.  The classical Higgs field can be
canonically normalized via
\begin{align}
\frac12 \gamma_{\phi\phi}(\phi) (\partial \phi)^2
= \frac{1}{2\Omega^2} \(1+\frac{6  \xi^2}{\mpl^2\Omega^2} \phi^2\) (\partial \phi)^2
= \frac12 (\partial h)^2 ,
\label{canonical}
\end{align}
with $ \Omega^2(\phi) = 1+ \xi \phi^2/\mpl^2$ evaluated on the
classical background.

The $v^2$-term in the Higgs potential plays no role during inflation,
and we set it to zero in the following.

%%%%%%%%%%%%%%%%%%%%%%%%%%%%%%%%%%%%%%%%%%%%%%%%%%%%%%

\subsection{UV completion and threshold corrections}
\label{s:UV}

We only consider higher order operators that change the inflationary
potential at loop level. This can be motivated independently in two
ways. Either assume that the UV completion respects an approximate
shift symmetry, which forbids the most dangerous operators that already
change the potential at tree level.  Or adopt a minimal approach to UV
corrections, only adding higher dimensional operators that are really
necessary for consistency of the theory. The result in both cases is
that the unknown UV physics only enters the inflationary potential via
their effect on the renormalization group equations and thus on
the running of the couplings. As we will show analytically in the next
section, the inflationary predictions are universal, and all
dependence on the running, and thus on the threshold corrections,
drops out.

Below we will motivate our choice of higher dimensional operators that
we add, and that we will use in our numerical computations. We would
like to stress, though, that this choice is not critical to our
results, and other parameterizations can be chosen and additional
corrections can be added.  As long as inflation is possible at all ---
no large corrections to the tree level potential and UV physics only
enters via the RGE equations--- the inflationary predictions are
robust.

It is well known that for a large non-minimal coupling $\xi \gg 1$, as
needed for Higgs inflation, unitarity of tree level scattering breaks
down well below the Planck scale. The unitarity cutoff --- the
momentum scale at which tree-level unitarity is violated --- depends
on the field value of the Higgs field, and is given by
\cite{bezrukov4,linde_higgs, cliffnew}:
\be
\Lambda \sim \left\{ \frac\mpl\xi, \phi , \frac\mpl{\sqrt{\xi}}\right \},
\label{unitary}
\ee
in respectively the small, mid, and large field regimes, defined via
\begin{align}
 {\rm small} \;{\rm field} : \; \phi  < \frac{\mpl}{\xi},\qquad
{\rm mid} \;{\rm field} : \; \frac{\mpl}{\xi}< \phi  < \frac{\mpl}{\sqrt{\xi}},\qquad
{\rm large} \;{\rm field} : \;\frac{\mpl}{\sqrt{\xi}}< \phi.
\label{regimes}
\end{align}
The field dependence of the cutoff may be understood from integrating
out heavy fields with a field dependent mass. In the case that the
cutoff signals the onset of strong dynamics, the field dependence also
may arise naturally.

Over the whole field region the typical energy in the Higgs potential
is below the cutoff $V(\phi)^{1/4} < \Lambda(\phi)$.  Nevertheless,
the field value during inflation exceeds the unitarity cutoff.
Following the usual effective field theory approach, and adding all
operators that respect the symmetries of theory, the model is
extremely sensitive to UV physics.  Indeed, all higher operators of
the form $(\Phi^\dagger \Phi)^{n+2}/\Lambda^n$ will completely spoil
the inflationary potential.  Also operators of the form
$(\Phi^\dagger \Phi)^{n}\O^4/\Lambda^n$ should be forbidden during
inflation; here $\O^4$ is a dimension four operator made up of
standard model fields, e.g. $\O^4 = F^{\mu\nu}F_{\mu\nu}$ and
$F_{\mu\nu}$ the SU(2) field strength tensor.  Indeed, during
inflation these operators will give rise to effective couplings that
are non-perturbatively large, and thus spoil the predictiveness of the
model. In this sense the situation in Higgs inflation is not different
from chaotic models of inflation.  In the latter case the cutoff is
the Planck scale and inflation takes place at superplanckian field
values, and thus also chaotic inflation is highly sensitive to
operators of the above form.

To avoid the dangerous operators discussed in the previous paragraph,
we assume that the UV completion respects the approximate shift
symmetry of the action in the inflationary regime, which is only
broken by a non-zero (but small) Higgs mass. This implies that at
dimension six, which are the leading corrections, only operators of
the form \cite{cliffnew}\footnote{In the small field regime $m_h^2
  \propto H^\dagger H$ and the operators \eref{L_threshold} reduce to
  the six dimensional operators listed in \cite{trott}. }
\be
\L \supset \sum_i c_i \frac{m_h^2}{\Lambda^2} \O_i^4
\label{L_threshold}
\ee
are allowed. Here $c_i$ are unknown Wilson coefficients, and the sum
is over all dimension four operators invariant under the SM
symmetries.  The cutoff is chosen as the field dependent unitarity
bound, which is an additional (but natural) assumption.  Using the
explicit form of the Higgs mass (see \eref{mass} below) and unitarity
bound \eref{unitary}, it can be seen that these operators are only
unsuppressed around the scale (we set $\mpl=1$ from now on)
$\phi \sim 1/\xi$. As a result, operators of the form
\eref{L_threshold} do not affect the tree level inflaton potential.
They can, however, affect the inflationary potential at the quantum
level, as these operators give corrections to the RGE equations
\cite{cliffnew,trott}. Running the SM couplings from the electroweak
scale, where they are measured, to the inflationary scale, one has to
pass the region where the normalization scale is $\mu \sim 1/\xi$ and
the threshold corrections --- if large enough --- cannot be neglected.

We can arrive at the same conclusion, i.e. threshold corrections that
are important only at $\phi \sim 1/\xi$, from a different perspective.
Namely: take the minimalistic approach to add new physics only when
really necessary for the consistency of the theory. For this purpose
we do not need, for the reasons we are going to explain in a moment,
higher dimensional operators that become important at the large field
values during inflation, only corrections around the scale
$\phi \sim 1/\xi$ are necessary.  First of all, although
$V^{1/4} < \Lambda$ at all field values, these scales become of the
same order at $\phi \sim 1/\xi$ and corrections to the Higgs
inflation action are unsuppressed. Secondly, the counterterms
introduced to absorb the UV divergencies of the quantum corrections
make a jump at the scale $\phi \sim 1/\xi$ \cite{damien2,shap}, which
signals that new physics should enter at this scale.  Let us explain
this second point in more detail.

The one-loop effective potential for the classical field $\phi$ is
given by the tree level potential plus the Coleman-Weinberg potential
\cite{CW}.  Explicitly
\be
V_{\rm eff} = V_{\rm tree} + \frac1{64\pi^2} \sum_i (-1)^{F_i} S_i
m_i^4(\phi) \[\ln \(\frac{m_i^2(\phi)}{\mu^2}\) -c_i \]
\label{VCW}
\ee
in the $\overline{\rm MS}$ renormalization scheme. Here $\mu$ is the
normalization scale, $F_i = 0\, (1)$ for a boson (fermion) field,
$S_i$ counts the degrees of freedom of each particle with mass $m_i$,
and $c_i = 3/2$ for fermions and scalars and $c_i =5/6$ for gauge
bosons. The gauge boson, top quark, Higgs and (three) Goldstone boson
(GB) masses are given (in Landau gauge) by \cite{damien2,shap}
\be
m_{A^i}^2 = \frac{g_i^2 \phi^2}{2\Omega^2}, \quad
m_t^2 = \frac{y_t^2 \phi^2}{2\Omega^2}, \quad
m_h^2 = 3 \lambda \phi^2 \frac{1+4\xi^2 \phi^2 -4 \xi^3
  \phi^4}{\Omega^4(1+6\xi^2 \phi^3)^2},\quad
m_{\theta^i}^2 = \frac{ \lambda \phi^2}{\Omega^4(1+6\xi^2\phi^2)},
\label{mass}
\ee
with $g_i = \{ g_2,\sqrt{g_1^2+g_2^2}\} $ for the $W$ and $Z$ bosons
with $g_1,g_2$ the hypercharge $U(1)_Y$ and and $SU(2)$ couplings
respectively, and $y_t$ the top Yukawa.  The loop contribution from
the gauge bosons and top quark has the same field dependence as the
tree level potential, and the divergencies can be absorbed in the
whole field range.  However, that is not the case for the Higgs and GB
masses.  The theory is not renormalizable over the full field range.
It has been shown in \cite{damien2} that nevertheless in the small,
mid and large field regimes \eref{regimes} a renormalizable EFT can be
constructed. That is, when the Lagrangian is expanded in a small
parameter that defines the given regime, all divergencies can be
absorbed order by order in a finite number of counter terms; no new
operators beyond those already present in the tree level Lagrangian
are needed. The EFTs are valid only within the given regime, and for
energies below the (field-dependent) unitarity cutoff \eref{unitary}.
The renormalization group equations in the small field regime are
those of the Standard Model.  The RGEs in the mid and large field
regimes are the same, and differ from the SM RGEs because of the
non-minimal coupling.  The EFTs need to be patched together at the
boundaries.  As is clear, at least at the border between the small and
mid field regime, which is at $\phi \sim 1/\xi$, threshold corrections
are needed as it is here that the RG equations change.

%%%%%%%%%%%%%%%%%%%%%%%%%%%%%%%%%%%%%%%%%%%%%%%%%%%%%%

\subsection{Renormalization prescription}
\label{s:mu}

Higgs inflation can be analyzed, and loop corrections can be
calculated in both the Jordan \eref{L_jordan} and Einstein frame
\eref{LE}. Even if the frames are merely related by a field
transformation it is not universally accepted that they describe the
same physics. The equivalence of the Jordan and Einstein frame
~\cite{hertzberg3,quiros1,quiros2} can be made explicit by rewriting
the action in terms of dimensionless quantities which are invariant
under a conformal transformation \cite{volpe,catena}. The equivalence
can also be checked on a case-by-case basis. For example, in
\cite{gong,chiba,kubota,janW1,janW2} it was shown that both frames
give the same result for the curvature perturbation during inflation,
\cite{christian} uses a covariant approach to show that both frames
gives the same (on-shell) effective action, and in \cite{moss} the
same covariant approach was used to show that the RGE equation for
$\xi$ is the same in both frames.  Finally, in \cite{damien2} it was
shown that the Coleman-Weinberg potential and the renormalization
procedure is one-to-one in both frames.

Despite all this there remains confusion in the literature on the
frame dependence of the results, and in particular on the choice of
renormalization scale.  Here we review that if the renormalization
prescription is done carefully no such ambiguity arises, more details
can be found in \cite{damien2}. Another way to arrive at the same
conclusion is to do the one-loop analysis using dimensionless quantities
invariant under a conformal transformation, the approach advocated in
\cite{volpe,catena} (which trivially corresponds to the Einstein frame
analysis).

To include the (one-loop) correction one could proceed in two
ways\footnote{In both cases we want to end up in the Einstein frame
  where slow roll inflation is most easily studied.}:
\begin{enumerate}
\item First go to the Einstein frame and then add the quantum correction
to $V_{E}$.
\begin{equation}
V_{E_1}=V_{E}^{(0)}(\phi)+V_{E}^{(1)}=\frac{V_{J}^{(0)}}{\Omega^{4}}+V_{E}^{(1)},
\end{equation}
where the superscript $(0)$ and $(1)$ refer to the tree level and
one-loop Coleman-Weinberg potential respectively.

\item Add the CW corrections to the Jordan frame potential
and only after transform to the Einstein frame
\end{enumerate}
\begin{equation}
V_{J}^{(0)}(\phi)+V_{J}^{(1)}\,\overset{E}{\longrightarrow}\, 
V_{E_{2}}=\frac{V_{J}^{(0)}(\phi)}{\Omega^{4}}+\frac{V_{J}^{(1)}}{\Omega^{4}}.
\end{equation}
As can be seen in the above equations, but this is general, all mass scales
are rescaled by the conformal transformation
\be
m_J \,\overset{E}{\longrightarrow}\, m_E = \frac{m_J}{\Omega}.
\label{mass_scaling}
\ee

If one does not consider the back reaction from gravity, following one
of the two paths leads to different results in the Higgs-gravity
sector. This is understandable since degrees of freedom
considered frozen in one frame are dynamical in the other and vice
versa.%
\footnote{For example, the Sasaki-Mukhanov variable is a different
  combination of the scalar metric degree of freedom and the Higgs in
  the two frames. To leading order in the slow roll approximation, one can
  treat gravity classically in the Einstein frame as the effects from
  back reaction are suppressed \cite{damien2}; however, this is not the case
  for the Jordan frame, and care should be taken when considering the
  Higgs and GB loops.}
Since the main contribution to the CW potential comes from the top quark
and gauge boson loops --- the Higgs and GB loops are suppressed --- we
do not have to worry about this.

Let's consider then the contribution of the top quark to the
Coleman-Weinberg potential in the Einstein frame, following procedure 1
\begin{equation}
V_{E_{1}}=\frac{(\lambda+\delta \lambda) \phi^{4}}{4\Omega^{4}}
+\frac1{8\pi^2}m_{t,E}^{4} \ln\left(\frac{\Lambda_{E}^{2}}{m_{t,E}^{2}}\right),
\end{equation}
where cutoff regularization has been used; $\delta \lambda$ is the
counter term, and the Einstein frame top mass has been defined in
\eref{mass}.  Choosing the counterterm
$\delta\lambda=-\frac{y^{4}}{(4\pi^2)}\ln\left(\frac{\Lambda_{E}^{2}}{\mu_{E}^{2}}\right)$
gives
\begin{equation}
V_{E_{1}}=\frac{\phi^{4}}{4\Omega^{4}}\left(\lambda
+\frac{y^{4}}{8\pi^2}\ln\left(\frac{\mu_{E}^{2}}{m_{t,E}^{2}}\right)\right).
\end{equation}
The log will be minimized for $\mu_E = m_{t,E}$; in the RG improved
action this will then minimize the error, see Appendix
\ref{A:RGimproved}.  This choice of renormalization scale is often
referred to as ``prescription 1''. For Higgs inflation it corresponds
to the usual prescription that the renormalization scale is chosen to be
the typical energy scale involved in the process.

Procedure 2 gives
\begin{equation}
V_{J}=\frac{(\lambda+\delta \lambda)
\phi^{4}}{4} +\frac1{8\pi^2}m_{t,J}^{4}\ln\left(\frac{\Lambda_{J}^{2}}{m_{t,J}^{2}}\right)
=\frac{\lambda\phi^{4}}{4}+\frac1{8\pi^2}m_{t,J}^{4}\ln\left(\frac{\mu_{J}^{2}}{m_{t,J}^{2}}\right)
\end{equation}
where $m_{t,J}= \Omega m_{t,E}$ is the top mass in the Jordan frame. In
the second expression we set the counterterm
$\delta\lambda=-\frac{y^{4}}{4\pi^2}\ln\left(\frac{\Lambda_{J}^{2}}{\mu_{J}^{2}}\right)$.
At this stage the log in the potential will be minimized for
$\mu_{J}(t)\sim m_{t,J}$. This choice is often referred to as
``prescription 2''.  However, this expression is still in Jordan frame
units. Expressing the renormalization scale in Planck units
\be
\frac{\mu_J}{m_{{\rm pl},J}} = \frac{m_{t,J}}{m_{{\rm pl},J}}
=\frac{m_{t,E}}{m_{{\rm pl},E}} =\frac{\mu_E}{m_{{\rm pl},E}},
\ee
we see it is exactly the same prescription as in the Einstein frame.
Here it should be noted that {\it all} mass scales, including the
Planck mass, cutoff scale and renormalization scale, are rescaled as
in \eref{mass_scaling} under a conformal transformation \footnote{The
  situation is completely analogue to going from a conformal FLRW
  metric to a Minkowksi metric by doing a conformal transformation
  with $\Omega = a(t)$ the scale factor. All masses are rescaled by
  the scale factor, {\it cf.} the physical momentum (the canonical
  momentum in the FLRW metric) and comoving momentum (the canonical
  momentum in the Minkowski metric) are related by
  $k_{\rm com} = k_{\rm phys}/a$. Spurious factors of $a(t)$ (the
  scale factor is by definition unobservable) only appear when
  comoving scales are erroneously compared to physical mass scales
  \cite{senatore}.}.  Finally, transforming the Jordan frame potential
to the Einstein frame we retrieve $V_{E_{2}}=V_{E_{1}}$.%
There is no ambiguity in the renormalization scale, which is correctly
given by prescription 1.
For definiteness, we will use in the next section
\be
\mu_E = \frac{\phi}{\Omega(\phi,\xi)}.
\label{mu1}
\ee

Although different renormalization prescriptions do not arise from
frame dependence, one could still argue that they encode different UV
completions of the theory. As discussed around \eref{VCW} no
counterterms can be defined that absorb the (subdominant) Higgs and
GB contributions over the whole field range.  Consider then  field
dependent counterterms of the form
$\delta\lambda=-\frac{y^{4}}{4\pi^2}\ln\left(\frac{\Lambda^{2}}{\mu^{2}}f(\phi)\right)$.
The Einstein frame potential becomes
\begin{equation}
V_{E}=\frac{\lambda\phi^{4}}{4\Omega^{4}}
 + \frac1{8\pi^2}m_{t}^{4}\ln\left(\frac{m_{t}^{2}}{\mu^{2}}f(\phi)\right).
\end{equation}
The choice $f(\phi)=\Omega^{2}$ corresponds to prescription 2, as
$\mu= f(\phi) m_t$ will minimize the log. Note however, that a field
dependent counterterm implies adding new operators to the
action. Indeed the above expression could only have come from a
potential of the form
\be
V_{\rm eff} = \frac{(\lambda+\delta \lambda)}{4\Omega^4} \phi^4 + \frac1{8\pi^2}
m_\tau^4 \ln\( \frac{\Lambda^2}{m_\tau^2} \) + \frac1{8\pi^2}
m_\tau^4 \ln\(f(\phi) \).
\label{V_f}
\ee
Therefore, a non-trivial choice of $f(\phi)$ implies that the
potential is already changed at the classical level! The simple form
of the action in the Jordan frame \eref{L_jordan}, with just a single
new parameter compared to the SM Lagrangian, can no longer be used as
a motivation for the model. Moreover, allowing for any UV completion
possible, i.e. for any choice of $f(\phi)$, all predictivity is lost
as literally any potential can be constructed. Fortunately, that is
not needed. The choice $f(\phi) =1$ is the natural choice as no new
operators and counterterms beyond those present in standard Higgs
inflation \eref{LE} are needed in the large field regime. As has been
shown in \cite{damien2} for $f(\phi)=1$ a renormalizable EFT can be
constructed in the small, mid and large field regime. On the
boundaries of these regimes, and as discussed in Section \ref{s:UV},
especially between the small and mid field regime, threshold
corrections are needed. But for consistency alone, adding new
corrections in the large field regime is not demanded.

Let us stress a crucial point about this way of parametrizing the
renormalization scale. We have already seen that the cutoff depends on
the Higgs vev. In the low field regime
$\Lambda_{{\rm low} \; {\rm field}}\equiv1/\xi$. If one is interested
in the RG flow at energy scales beyond $1/\xi$ it might seem it is not
possible to say anything without knowing exactly the form of the UV
completion.  Consider the analogy with the Fermi effective theory of
beta decay. At energy values below the $W$-boson mass the Fermi
effective action can be used to compute the beta functions
etc. However, for energies above the cutoff ($W$ mass) the knowledge
of the full electroweak Lagrangian is needed. The situation here is
considerably different. The "prescription 1" choice of the
renormalization scale automatically takes in account that when
$\mu(\phi) > \Lambda_{{\rm low} \; {\rm field}}$ the field is no
longer in the low field region and the unitary bound is still larger
than $\mu$.\footnote{Prescription 1 gives $\mu(\phi) < \Lambda(\phi)$
  for all field values.  For prescription 2, on the other hand
  $\mu(\phi) > \Lambda(\phi)$ for large field values, and the RGE
  evolution can no longer be described in the EFT setting, the full UV
  completion is required.}  Thus the full form of the UV completion is
not needed and one can consistently parametrize it with a series of
higher order operators suppressed by the scale given by the field
dependent cutoff.\footnote{If a constant cutoff
  $\Lambda_{{\rm low} \; {\rm field}}$ is chosen you would need for
  instance an extra degree of freedom to restore the unitarity of the
  model till the Planck scale \cite{giudice,mirage}. These UV
  completions modify the inflaton potential already at tree level, and
  thus they are different from the ones discussed in this paper.} From the physical point
of view the difference between the two situations can be understood
from the fact that here the increase in energy is due to a
displacement of the Higgs vev.

%%%%%%%%%%%%%%%%%%%%%%%%%%%%%%%%%%%%%%%%%%%%%%%%%%%%%%

\subsection{Renormalization group equations}
\label{s:RGE}

In the next section we will calculate the inflationary observables
taking into account the running of all couplings. In particular, we
consider the RG improved action \eref{LE} with potential and field
space metric (from now on we set $V_{\rm eff}\equiv V$)
\be
V = \frac{\lambda(t) \phi^4}{4(1+\xi(t) \phi^2)^2}, \qquad
\gamma=\frac{1+\xi(t) \phi^2(1+6\xi(t))}{(1+\xi(t) \phi^2)^2},
\label{L_running}
\ee
with 
\be
t = \ln(\mu/m_{\rm t})
\label{t_def}
\ee
and $m_t$ the EW scale top mass. In addition, the renormalization
scale \eref{mu1} depends on the running couplings
\be
\mu = \frac{\phi}{\sqrt{(1+\xi(t) \phi^2)}}.
\label{mu2}
\ee
For more details see the discussion in appendix \ref{A:RGimproved}.

The running of the couplings is governed by the RG equations. In the
small field regime these are just the SM RGEs, we use the two-loop
result and the EW boundary conditions defined in \cite{kyle,RGE}. The
one-loop RGEs for the mid and large field regime have been derived by
different groups \cite{bezrukov_loop,
  bezrukov3, wilczek, barvinsky, barvinsky2,
  barvinsky3,damien,kyle,GMP}, with small differences.  We use the recent
result of \cite{damien2}. Our main conclusions will not depend on this
choice, only the exact numerical values of parameters might differ
slightly. We set the boundary condition for $\xi_0$ at the boundary of
the mid-field regime $\xi(1/\xi) =\xi_0$.

In \cite{cliffnew,trott} the corrections to the $\beta$ functions due to the
higher dimensional operators \eref{L_threshold} were calculated. The
corrections depend on unknown Wilson coefficients $c_i$, the  Higgs
mass which is given in \eref{mass}, and the cutoff scale that we
choose 
\be
\Lambda =  \frac{(1+\xi(t)^{2} \phi^2)}{\xi(t)^2(1+\xi(t) \phi^2)},
\ee
which reduces to the unitarity cutoff in the three regimes
\eref{unitary}.  Since the operators are peaked at $1/\xi$,
only around this scale the corrections to the running are
appreciable. For inflationary purposes the effect is that threshold
corrections may give a ``kick'' to $\lambda$, i.e. change
$\lambda(\mu \sim 1/\xi)$ by some amount compared to the SM running.
Since $\lambda \ll 1$ this kick may be appreciable for Wilson
coefficients $c_i \sim \O(10)$ (such that the threshold and SM
contribution to the RGEs are of comparable size
$\delta \beta \sim \beta_{\rm SM}$ at the scale $\mu = 1/\xi$). The
relative kick to other SM parameters is very small.  For our numerical
results we choose the Wilson coefficients, defined in the appendix B
of \cite{cliffnew}, randomly in the interval
\be
c_i = {\rm Random}[-c^{\rm max}, c^{\rm max}].
\label{c_max}
\ee
In our numerics, we choose boundary conditions at the EW scale,
$\xi_0$ at the intermediate scale, and a set of Wilson coefficients
$c_i$, and then run all couplings to the large field regime.  We then
determine $t_{\rm end}$ and $t_\star$, i.e. the normalization scale
\eref{t_def} at the end and $N_\star$ efolds before the end of
inflation, and finally the power spectrum for the perturbations. We
reiterate this procedure, adjusting the value of $\xi_0$ till the
right COBE normalization \eref{xival} is obtained. It may happen
that for some or all $\xi_0$-values inflation with more than $N_\star$
efolds is impossible. For definiteness, we take $N_\star =60$.  In the
next section we discuss the calculation of the perturbations during
inflation in detail.

%%%%%%%%%%%%%%%%%%%%%%%%%%%%%%%%%%%%%%%%%%%%%%%%%%%%%%
%%%%%%%%%%%%%%%%%%%%%%%%%%%%%%%%%%%%%%%%%%%%%%%%%%%%%%
%%%%%%%%%%%%%%%%%%%%%%%%%%%%%%%%%%%%%%%%%%%%%%%%%%%%%%

\section{Inflation}
\label{s:inflation}

In this section we compute the spectral index and tensor-to-scalar
ratio taking into account the running of the couplings. We will show
analytically in the next subsection that for inflation on a flat
plateau, as it is usually assumed in Higgs inflation, all dependence on
the beta-functions drops out, and the inflationary observables are the
same as for the classical potential. With running included, it is
possible for a limited range of parameters that the potential develops
a maximum. As discussed in subsection \ref{s:max} for hilltop
inflation, i.e. inflation near the maximum, the results depend
sensitively on the running, and thus on the UV completion (entering
via the beta-functions). We present numerical results for this case.

%%%%%%%%%%%%%%%%%%%%%%%%%%%%%%%%%%%%%%%%%%%%%%%%%%%%%%
\subsection{Inflation on the flat plateau}
\label{s:flat}

Higgs inflation takes place in the large field regime \eref{regimes}, where
the action can be expanded in the small parameter
\be
\delta={1}/{(\xi \phi^2)} \ll 1.
\label{delta}
\ee
As follows from (\ref{sr},~\ref{efolds}) below, the $\delta$-expansion
is equivalent to an expansion in slow roll parameters, and is also
equivalent to an $1/N_\star$ expansion.

In order to include the effects of running
couplings on the inflationary observables we work with the
renormalization group improved action. The potential and field space
metric \eref{L_running} can be rewritten as
\be
V =
\frac{\lambda(t)}{4\xi(t)^2} \frac1{(1+ \delta(t))^2}, \qquad
\gamma_{\phi\phi} = \frac{\delta(t)(1+\delta(t) +6 \xi(t))}{(1+\delta(t))^2},
\qquad
\delta(t)= \frac{1}{\xi(t) \phi^2} ,
\ee
with $t = \ln(\mu/m_t)$, $m_t$ the EW scale top mass, 
and $\mu$ the
renormalization scale \eref{mu2}
\be
\mu = \frac{1}{\sqrt{\xi(t)(1+\delta(t))}},
\label{mu3}
\ee
which is proportional to the top and gauge boson
mass. This choice minimizes the logs in the Coleman-Weinberg
expansion, as already discussed in section \ref{s:mu}. For $\delta \ll 1$ the
potential reduces to a constant plus (exponentially) small
corrections, and inflation takes place on a flat plateau. The running
of the couplings may slightly tilt the plateau, and thus change the
expressions for the observables.

To calculate the slow roll parameters the first and second derivatives
of the potential with respect to the canonically normalized field $h$,
defined in \eref{canonical}, are needed. Let's start with the slope
first. Using the chain rule gives
\be
V_h = \frac{1}{\sqrt{\gamma_{\phi\phi}}}\(
\frac{\partial V}{\partial \phi}+\frac{\partial V}{\partial 
      \lambda} \lambda_{\phi}+\frac{\partial V}{\partial \xi
      }\xi_{\phi}\),
\ee
with 
\be\label{derphi}
\lambda_{\phi} =
 %\frac{\partial  \lambda}{\partial t}\frac{\partial t}{\partial \phi}=
 \beta_{ \lambda} \frac{dt}{d\phi},
\qquad
\xi_{\phi} =
% \frac{\partial  \xi}{\partial t}\frac{\partial t}{\partial \phi}=
\beta_{ \xi} \frac{dt}{d\phi}, \qquad
\frac{dt}{d\phi} =
\frac{\delta^{3/2} \xi^{1/2}}{1+ \delta +\frac{\beta_\xi}{2\xi}},
\ee
where we used the definitions
$\beta_\lambda = {\partial \lambda}/{\partial t}$ and
$\beta_\xi = {\partial \xi}/{\partial t}$. The last expresson follows
from the explicit form of normalization scale \eref{mu3}.  Putting
it all together gives
\be
\frac{V_h}{V} = \sqrt{\frac83} \frac{\delta(1+\delta)}{\sqrt{1+
    \frac{(1+\delta)}{6\xi}}} \frac{ (1 +
  \frac{\beta_\lambda}{4\lambda})}{(1+ \delta +
  \frac{\beta_\xi}{2\xi})}.
\label{dV}
\ee 
This result is still exact, no $\delta$-expansion or other
approximation has been done.  In a similar way the 2nd derivative of
the potential can be computed.  The slow roll parameters become
\be
\eps \equiv \frac12 \( \frac{V_h}{V} \)^2
 =\frac43 \delta^2 F^2 \(1+ \frac1{6\xi}\)
     +  \O(\delta^3) ,\qquad
\eta \equiv \frac{V_{hh}}{V} 
= -\frac43 \delta F
   +    \O(\delta^2),
\label{sr}
\ee
with 
\be
F = \frac{
(1 + \frac14 \frac{\beta_{\lambda}}{\lambda})}{ (1+  \frac12
\frac{\beta_\xi}{\xi} ) (1 + \frac1{6\xi})}.
\label{F}
\ee
Turning off the running of the couplings
$\beta_{ \lambda} = \beta_\xi =0$, we retrieve the standard classical
results (which are often expressed in the large $\xi$ limit where
$F=1 + \O(1/\xi)$).

The RGEs for SM Higgs inflation have been calculated in the
literature.  In the large field regime they differ from the SM ones;
we quote the recent results \cite{damien2}
\be
\frac{\beta_{\lambda}}{\lambda} = \frac1{8\pi^2} \(
\frac{3g^4-y^4}{\lambda} - 2y^2\), \qquad
\frac{\beta_{\xi}}{\xi}  = \frac1{8\pi^2} y^2.
%\frac{\beta_{\bar \lambda}}{\bar \lambda} = \frac1{8\pi^2} \(
%\frac{3g^4-y^4}{\lambda}  \).
\label{betafunctions}
\ee
Note that in the inflationary regime the contribution from threshold
corrections to the beta functions can be neglected since for our
choice their contribution is important only around the scale
$\mu =1/\xi$.  The main point is that $\beta_\xi/\xi < 1$ is always
perturbatively small and the denominator of $F$ is always positive.
The top contribution dominates and $\beta_\lambda < 0$ at the
inflationary scale. This means that $F$ can go through zero and become
large and negative in the $\lambda \to 0$ limit.  When
\be
F = 0 \quad \Leftrightarrow \quad \lambda_{\rm max} = -
\frac{\beta_\lambda}{4},
\label{lambda_max}
\ee
to the lowest order in the $\delta$-expansion the slow roll
parameters vanish.  As can be seen from \eref{dV} this corresponds to
an extremum of the potential, and $\eps =0$ at all orders. For SM
Higgs inflation $\lambda_{\rm max} \sim 5 \times 10^{-5}$. For
energies well below the Planck scale the quartic coupling
$\lambda(t)$ is a monotonically decreasing function, and there is at
most one extremum which is a maximum as~\footnote{Close to the Planck
  scale 
%the $U(1)$ coupling increases rapidly as it nears the Landau
%  pole. 
there is the possibility of a second extremum, a minimum,
  in the potential. This opens the possibility for inflation near an
  inflection point. We comment on this in section \ref{s:max}.}
\be
\eta \big |_{\lambda = \lambda_{\rm max}} = - \frac{8}{3}\frac{
    \delta^2(1+\delta)^2 (1 +\frac{\beta_\lambda'}{ 4\beta_\lambda})}{(1 + \frac{1+\delta}{6\xi}) (1+ \delta
    +\frac{\beta_\xi}{2\xi})^2} < 0.
\label{sr_crit}
\ee
If the potential develops a maximum in the inflationary regime, 60
efolds of inflation may still occur if the potential near the maximum
is flat enough. We will refer to this possibility as ``hilltop
inflation''. To end up in the electroweak vacuum of the Higgs
potential, this should happen for field values $\phi < \phi_{\rm max}$
where $\phi_{\rm max}$ is the field value at the maximum (this
possibility thus constrains the initial field values); this corresponds
to the region where $F>0$ is positive.  Note that the $\delta$-expansion breaks
down close to the maximum, when 
\be
F \approx \(1 +\frac14 \frac{\beta_\lambda}{\lambda}\)
\sim \delta  \(1 +\frac{\beta_\lambda'}{ 4\beta_\lambda}\),
\label{Fcrit}
\ee
and the first order term of $\eta$ in \eref{sr} becomes comparable to
the $\delta^{2}$ term in \eref{sr_crit}  (we took the $\xi \gg 1$ limit).

Introduce the notation that the subscript $\star$ denotes the value of the
parameters when observable scales leave the horizon, $N_\star$ number
of efolds before the end of inflation. We distinguish three
possibilities.
\begin{enumerate}
\item If $F_\star \gtrsim \delta_\star$ inflation takes place on a flat
  plateau, and there is no maximum.\footnote{Note that
    $\lim_{\phi \to \infty} \mu^2 = 1/\xi$ approaches a constant, and
    the running comes to a halt. For $F_\star > \delta_\star$ the
    asymptotic value of $\lim_{\phi\to \infty} \lambda(t)$ exceeds the
    critical value \eref{lambda_max}.}  This is the case for  coupling
  values $\lambda_\star \gtrsim 5 \times 10^{-5}$.
\item If $F_\star \lesssim \delta_\star$ there is a maximum in the potential.
  If the maximum is flat enough, hilltop inflation takes place
  close to the maximum at field values $\phi < \phi_{\rm max}$.
 This is the case for  coupling
  values $\lambda_\star \sim 5 \times 10^{-5}$.
\item 
The potential near the maximum is too steep to support $N_\star =60$
efolds of inflation.
\end{enumerate}
In this section we will discuss case 1, inflation on the flat
plateau. The discussion of case 2, hilltop inflation, is postponed
till the next section.\footnote{It may happen that threshold
  corrections kill Higgs inflation, in that the corrections to the
  RGEs bring the model from case 1 to case 3.  The statement we want
  to make in this paper is that when inflation happens, the
  predictions are robust and insensitive to UV corrections (except for
  some possible fine-tuned parameters that allow for
  hilltop/inflection point inflation). }  The value
$F_\star \sim \delta_\star$ divides the two regimes, as follows from
\eref{Fcrit}; this is in agreement with our numerical results, which
are presented in \ref{s:max}.  The slow roll parameters \eref{sr} are
affected by the running of the couplings, and corrections may become
sizeable for small $\lambda$. However, to calculate the inflationary
observables, the slow roll parameters are to be evaluated at the field
value $\phi_\star$ at which the observable scales leave the
horizon. This field value also gets corrected by the running, and as
we will show now, these corrections exactly cancel, such that the
inflationary predictions are to leading order in the
$\delta$-expansion not affected by the running of the couplings.

Let's thus compute the number of efolds $N_\star$ before the end of
inflation, which is given by
\be
N_\star \simeq\int^{h_\star} \dd h 
\frac1{\sqrt{2\eps}} \simeq \frac{ \sqrt{3}}
{|F_{\star}| \sqrt{ 8(1 + \frac1{6\xi_{\star}})}} 
 \int^{h_\star} \dd h \, \delta^{-1}
=\frac1{\delta_\star|F_{\star}|} \frac34.
\label{efolds}
\ee
On the flat plateau $F > 0$ and we can drop the absolute signs.  Here
we assumed that $F$ and $\xi$ is to first approximation
field-independent and we have taken it out of the integral. This gives
the leading term in the $\delta$-expansion, as we now quickly explain.
The efolds integral can be rewritten as follows
\be
N_\star = 
\frac3{2}\int^{\phi^\star}\dd\phi\,\frac{\xi}{|F|}\phi +O(\sqrt{\delta})
=\frac{3}{2}\int^{\phi_\star} \dd
\phi\,\phi\left(D_{\star}+\frac{dD}{d\phi}\big|_{\phi=\phi_\star}(\phi-\phi_\star)+..\right).
\label{eq:exp}
\ee
In the first step we used the field space metric to express the
integral in terms of the (non-canonical) field $\phi$
\eref{canonical}, in the second step we defined
$ D(\beta_{i},\lambda,\xi) \equiv \xi |F|^{-1}$ and expanded the
integrand around $\phi_\star$. The first term in the expansion is the only
one considered in \eref{efolds}. It gives a contribution of the form
\begin{equation}
\int^\star\dd
\phi\,\phi D_\star\propto\phi_\star^{2}\propto O(\delta_\star^{-1}).
\label{eq:ll}
\end{equation}
The second term in the expansion takes the form ($f_{i}\equiv \{\lambda,\xi,\beta_{\lambda},\beta_{\xi}\}$),
\begin{equation}
\int^{\phi_\star}\dd
\phi\,\frac{dD}{d\phi}\bigg|_\star\phi(\phi-\phi_\star)
=\int^{\star}\dd
\phi\,\phi(\phi-\phi_\star)\left(\frac{dD}{df_{i}}
\frac{df_{i}}{dt}\frac{dt}{d\phi}\right)\bigg|_\star
\propto\phi_\star^{3}\delta_\star^{\frac{3}{2}} \propto
O(\delta_\star^{0}),
\end{equation}
where we used $dt/d\phi|_\star \propto \delta_\star^{\frac{3}{2}}$
from \eref{derphi}.  It is higher order in the $\delta$ expansion and
can be neglected.  We also neglected the lower bound of the integral;
this correction is likewise higher order in
$1/N_\star \sim \delta_\star$.
Using \eref{efolds} from the COBE normalization we get \cite{Planck}
\be
\(\frac{V}{\eps}\)_\star =
\frac43 \frac{\lambda}{\xi^2 N_\star^2}
= (0.027)^4 
\qquad \Rightarrow \qquad
\frac{\xi(t_\star)}{ \sqrt{\lambda(t_\star)} }= 5 \times 10^4 .
\label{xival}
\ee
Plugging  \eref{efolds} in the expressions \eref{sr} gives the spectral index and tensor-to-scalar ratio
\be
n_s =1 + 2 \eta + \O(\delta^2)
% = 1-\frac83 \delta_\star F + \O(\delta^2)
=1-\frac2{N_\star} + \O(\delta^2) 
\ee
and
\be
r = 16 \eps
%= 16 \times \frac43 \delta_\star^2 F^2 (1+ \frac1{6\xi})     +  \O(\delta^3) 
= \frac{12}{N_\star^2} \(1 + \frac1{6\xi}\) +  \O(\delta^3) 
\ee
The COBE normalization can always be fit by choosing the non-minimal
coupling appropriately.  All parameters in the model are then fixed.
For the large non-minimal couplings needed ($\xi_\star > 10^2$), the
spectral index and tensor-to-scalar ratio only depend on the number of
efolds. All dependence on the beta-function has cancelled in the final
expression, and the results are identical to those for classical Higgs
inflation. This means that the results for plateau inflation are very
robust: they are independent from the running, and thus insensitive to
UV physics that change the running, and also independent of the
electroweak boundary conditions on the couplings.

At next order in the $\delta$ expansion the beta-function do enter,
see appendix \ref{A:perturb}, but this is too small an effect to be
measured.

%%%%%%%%%%%%%%%%%%%%%%%%%%%%%%%%%%%%%%%%%%%%%%%%
\subsection{Inflation near the maximum}
\label{s:max}

As discussed previously, for $F_\star > \delta_\star$ inflation takes
place on the flat part of the potential and the inflationary
observables are insensitive to the running of the couplings, to first
order they depend only on the number of efolds. Here we discuss the
case $F_\star < \delta_\star$.  When $F=0$ the potential develops a
maximum. Requiring the Higgs field to end up in the electroweak
vacuum, inflation should take place at field values before the
maximum, where $F >0$. We expect Hilltop inflation to be highly
sensitive to the form of the potential and thus to the details of EW
boundary conditions and to threshold corrections. Unfortunately,
because of this sensitivity, it is hard to obtain analytical
expressions, and we will only present numerical results.  We point out
that even though hilltop inflation is sensitive to the UV completion,
it only occurs for very fine-tuned boundary conditions. Thus the
numerical results presented in this section confirm our statement that
the predictions for Higgs inflation are remarkably robust, and they
verify the analytical result of section \ref{s:flat}.  For more
details on our numerical implementation see section \ref{s:action} and in
particular \ref{s:RGE}.

\begin{figure}[t!]
\centering
\includegraphics[width=0.4\linewidth]{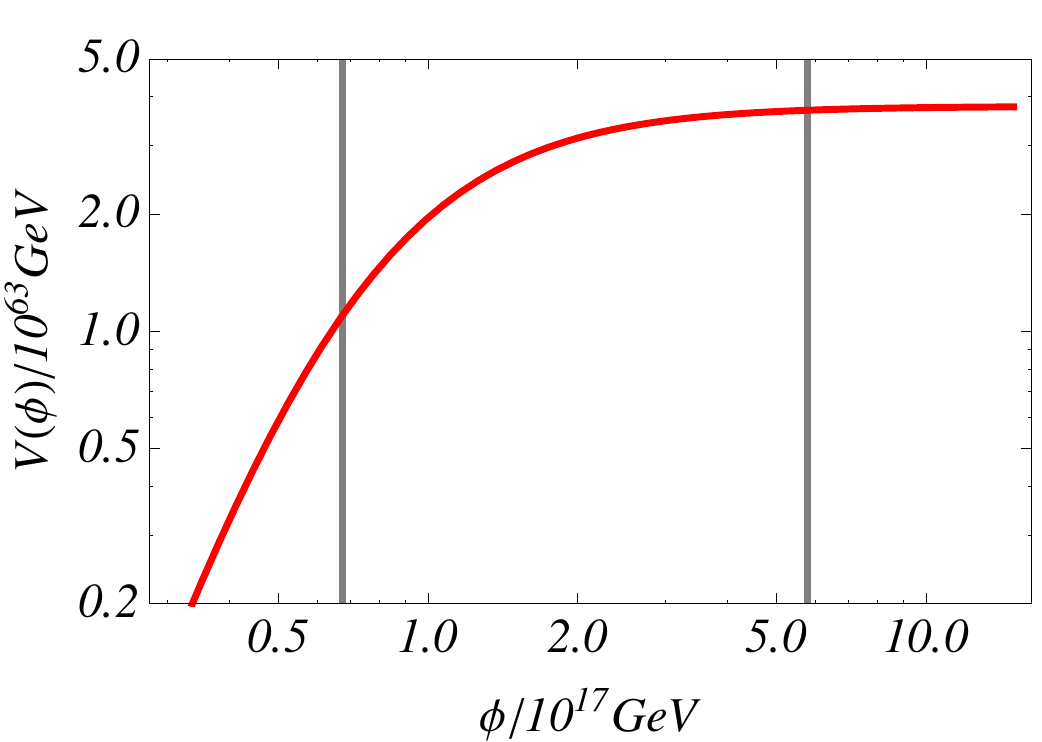}~~~
\includegraphics[width=0.42\linewidth]{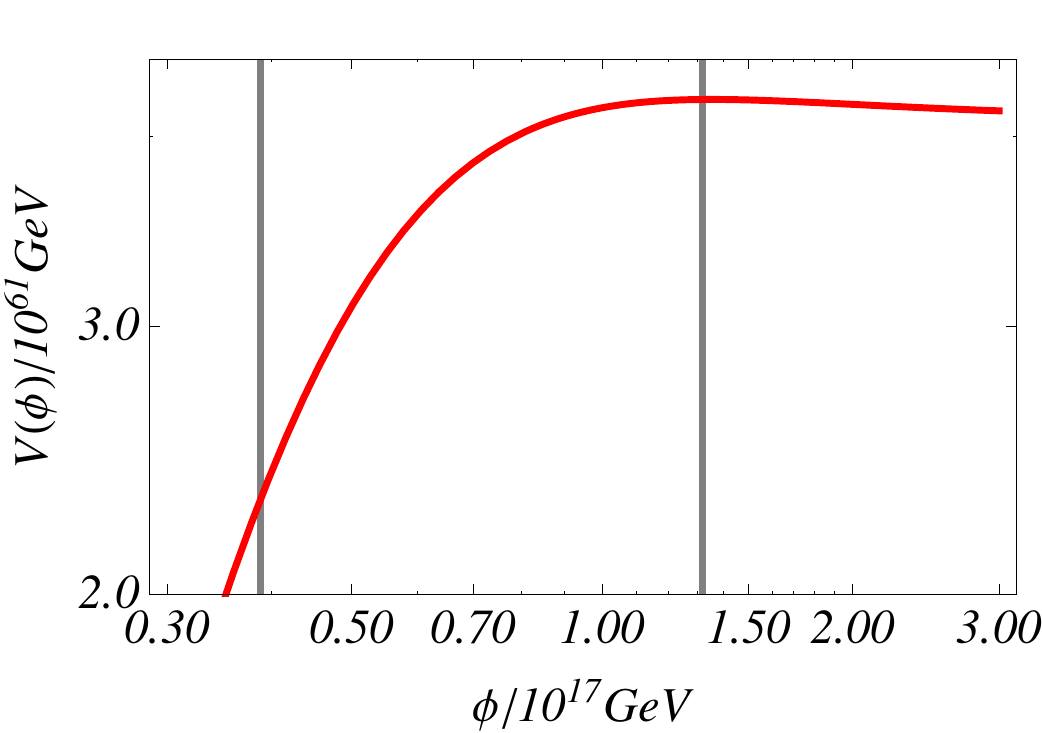}\\[0.3cm]
\includegraphics[width=0.4\linewidth]{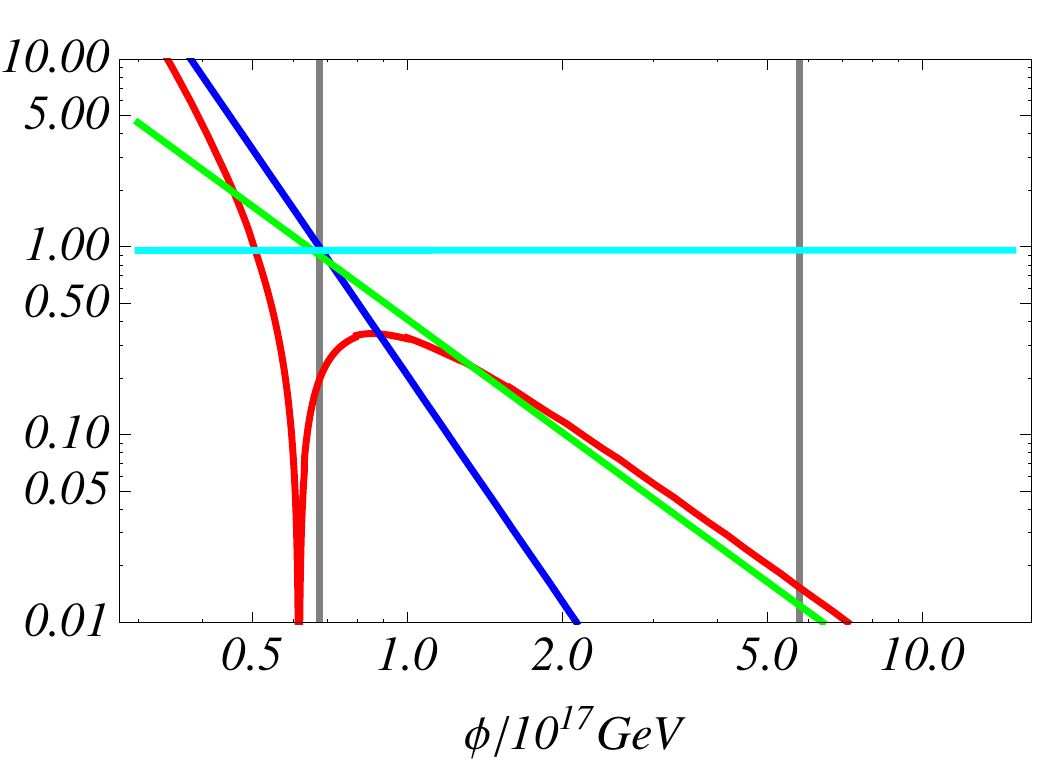}
\hspace{0.5cm}
\includegraphics[width=0.4\linewidth]{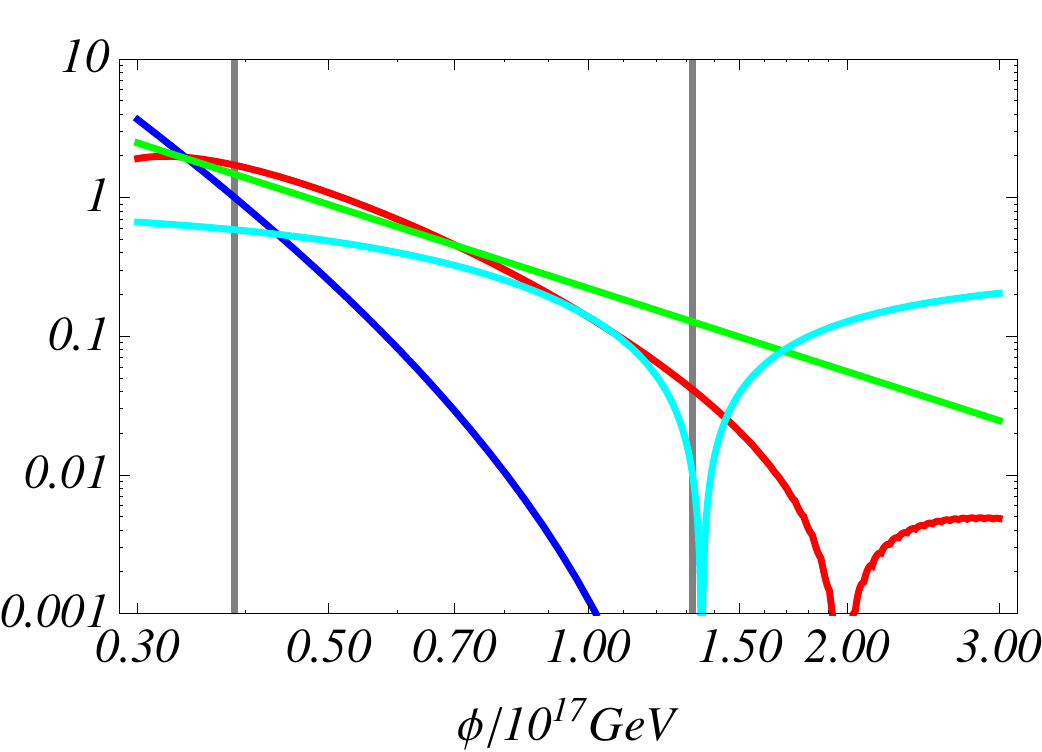}
\caption{Top left the potential $V(\phi)$ for inflation with
  $m_t =171$GeV and $m_h = 125.5$ GeV; the grey vertical lines
  correspond to $\phi_\star$ and $\phi_{\rm end}$ respectively
  (i.e. the beginning and end of inflation). Bottom left shows
  $|\eta| $ (red), $\eps$ (blue), $\delta$ (green) and $F$ (cyan)
  respectively for the same top mass. Right top and bottom, same plots
  but for $m_t = 171$ GeV and $m_h = 125.245$ GeV.\vspace{1.5cm}}
\label{F:V_SM}
\end{figure}

Let's start by considering just the SM running, and turn off all
threshold corrections.  We can tune $F$ small by adjusting the
boundary conditions at the electroweak scale. We choose to decrease
the Higgs mass, while keeping the top mass and gauge couplings
fixed.\footnote{We choose $m_t =171$GeV which is about $2\sigma$ below
  its central value, to avoid that the quartic coupling becomes
  negative before inflation.}  Our results are summarized in Table
\ref{Table:higgs}.  They agree with the discussion above.  For large
enough Higgs mass values, inflation takes place on the flat plateau,
and $n_s$ and $r$ are independent on the running.  In some fine-tuned
range of Higgs mass values, inflation can happen near a maximum; in
this case the inflationary results depend sensitively on the EW
boundary conditions.  For an even smaller Higgs mass the maximum is
too steep and 60 efolds of inflation is not possible.  In
Fig.~\ref{F:V_SM} we show an example potential for inflation on the
plateau and for inflation near the maximum\footnote{For these plots we
  numerically inverted $t(\phi)$. This inversion is not needed to
  calculate $n_s$ and $r$, which is done with $t$ as the clock
  variable.}, the parameters corresponding to the first and last line
of Table \ref{Table:higgs}.  Our numerical results agree with similar
studies in the literature \cite{bezrukov3,kyle}.

\begin{figure}[t!]
\centering
\includegraphics[width=0.48\linewidth]{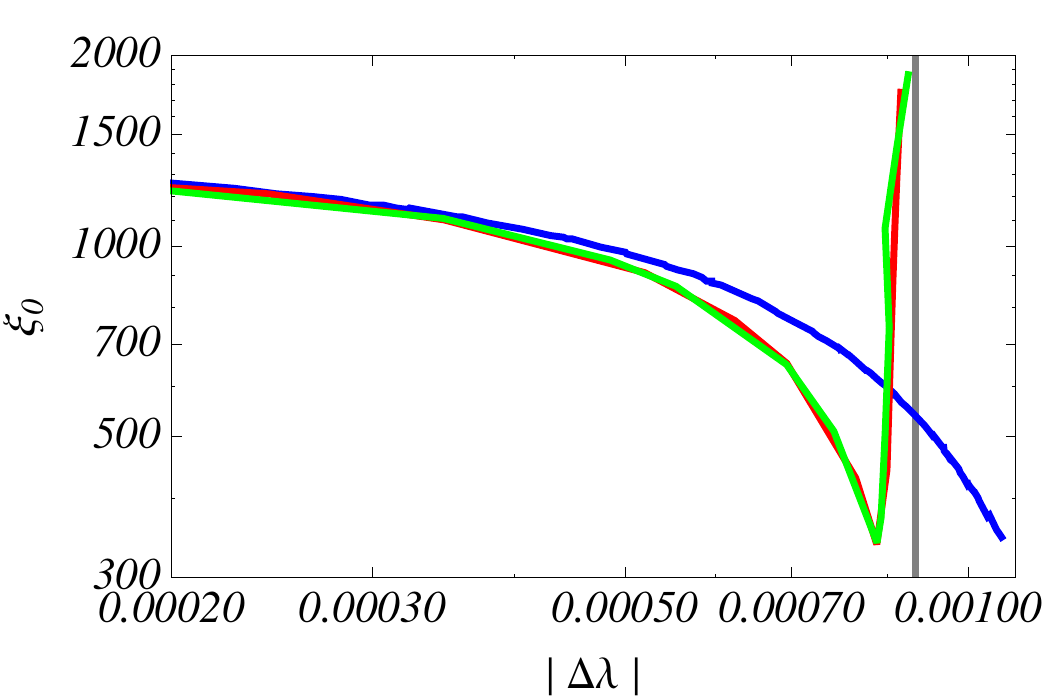}~~~
\caption{The boundary value $\xi_0$ for the non-minimal coupling
  vs. the kick $|\Delta \lambda|$ for SM running (green/red) and SM
  running with threshold corrections (blue).  In the former case the
  kick is from changing the top/Higgs mass at the electroweak scale,
  whereas in the latter it is due to the threshold corrections. The
  kick is downwards $\Delta \lambda<0$. The vertical grey line
  corresponds to the value $\lambda_*^{\rm SM} + \Delta \lambda =0$.}
\label{F:kick_xi}
\end{figure}

\begin{table}[h!]
  \centering
$
  \begin{array}{|l|c|c|r|r|c|c|c|}
\hline
    m_h({\rm GeV}) & h_\star  & \lambda_\star & \xi_\star\;\;\; & F_\star\;\;\; & \delta_\star & n_s &
   r \\
    \hline
127 &  0.15 & 6.3 \times 10^{-3}  & 3863 & 0.99 & 0.01&
0.968 & 3.0\times10^{-3} \\
\hline
126 &  0.18 & 2.7 \times 10^{-3}  & 2505 & 0.98 & 0.01 &
0.968 & 3.0\times10^{-3} \\
\hline
\rowcolor{yellow}
125.5 &  0.24 & 9.0 \times 10^{-4}  & 1451 & 0.96 & 0.01&
0.968 & 3.0\times10^{-3} \\
\hline
125.3 &  0.33 & 1.9 \times 10^{-4}  & 667 & 0.84 &0.01&
0.968 & 2.9\times10^{-3} \\
\hline
125.26 &  0.34 & 4.2 \times 10^{-5}  & 344 & 0.42 &0.03&
0.970& 2.4\times10^{-3} \\
\hline
125.255 &  0.20 & 3.1 \times 10^{-5}  & 451 & 0.12 & 0.06&
0.968 & 9.5\times10^{-4} \\
\hline\hline
125.253 &  0.13 & 3.3 \times 10^{-5}  & 730 & 0.05 & 0.08&
0.958 & 3.7\times10^{-4} \\
\hline
125.25&  0.09 & 3.7 \times 10^{-5}  & 1314 & 0.07 &0.12 &
0.941 & 1.2\times10^{-4} \\
\hline
125.245 &  0.05 & 4.3 \times 10^{-5}  & 2678 & 0.01 &0.12 &
0.917 & 3.4\times10^{-5} \\
\hline
  \end{array}
$
\caption{Inflationary parameters for different Higgs mass
  while $m_t =171 {\rm GeV}$ is kept fixed, in the absence of threshold
  corrections. Above the double line the potential has a flat
  plateau, below the line the potential develops a maximum. For $m_h <
125.245$ no inflationary solutions with $N_\star =60$ efolds exists.}
\label{Table:higgs}
\end{table}
\begin{figure}[t!]
\centering
\includegraphics[width=0.48\linewidth]{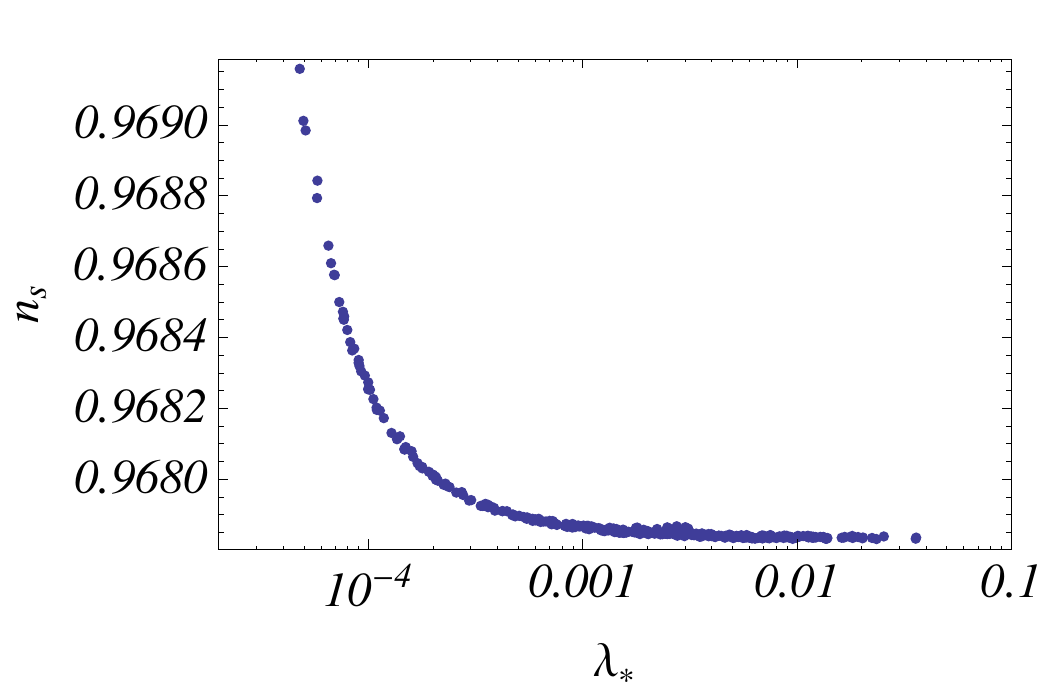}~~~
\includegraphics[width=0.48\linewidth]{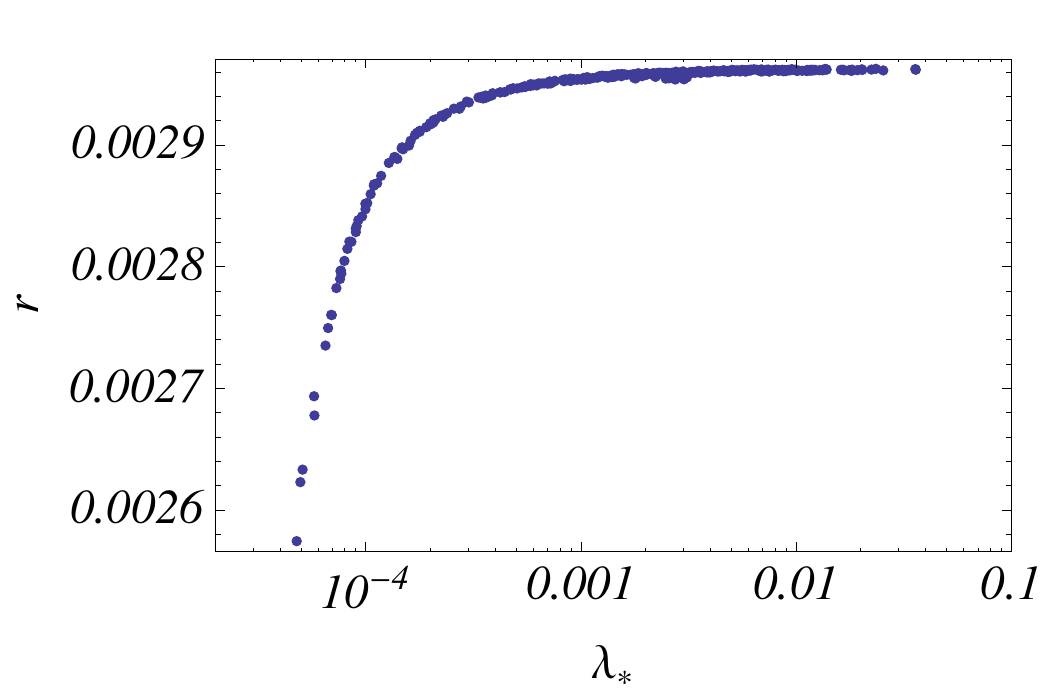}\\[0.3cm]
\includegraphics[width=0.48\linewidth]{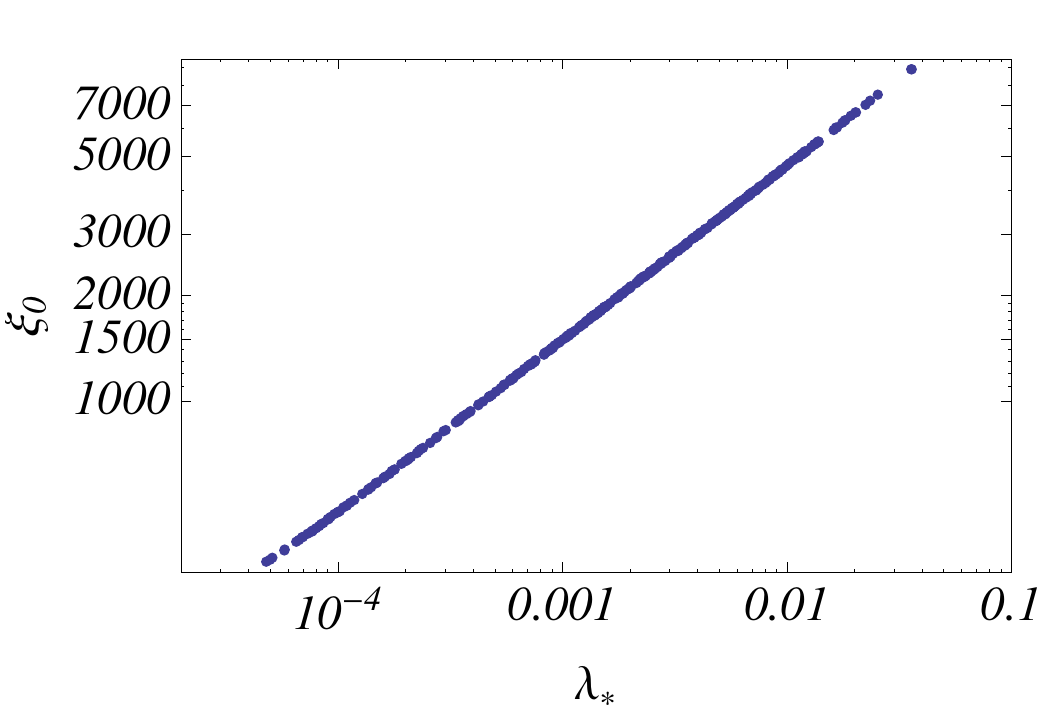}
\includegraphics[width=0.48\linewidth]{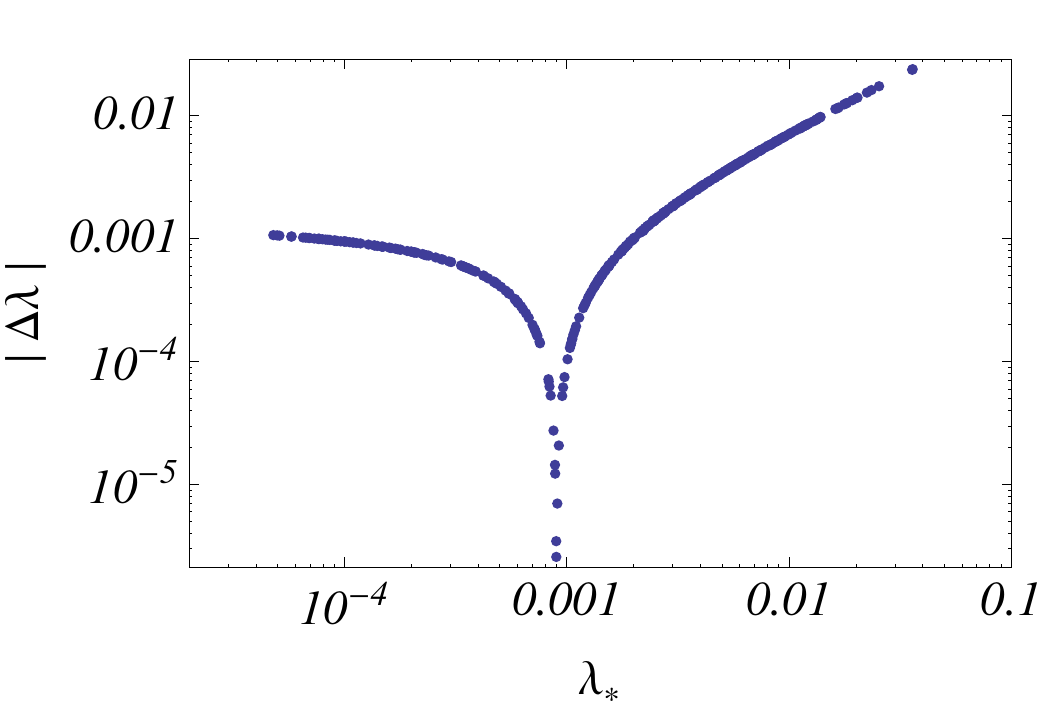}
\caption{The spread in spectral index $n_s$, tensor-to-scalar ratio $r$, $\xi_0$
and $\Delta \lambda$ as a function of $\lambda_\star$ for the 382 succesfull
models with threshold corrections $c_i ={\rm Random}[-10,10]$.}
\label{F:scatter}
\end{figure} 
$\qquad$
Now turn on the threshold corrections.  We choose $m_t = 171$ GeV,
$m_h = 125.5$ GeV and did 500 simulations with Wilson coefficients
randomly chosen between $c_i ={\rm Random}[-10,10]$.  We found 382
times that inflation takes place on the flat plateau, and the other
118 times there was no inflationary solution. Hilltop inflation does
not happen. The spread in spectral index, tensor-to-scalar ratio,
$\xi_0$ and kick $\Delta \lambda$ for the successful models are shown
in Fig.~\ref{F:scatter}.  The kick in $\lambda$ is defined with
respect to the reference set-up without threshold corrections,
corresponding to the highlighted line in Table \ref{Table:higgs}. Define
$(\lambda_\star^{\rm SM}, t_\star^{\rm SM},\xi_0^{\rm SM}) = (9.0
\times 10^{-4},\, 33.55,\,1417)$ for this
model, with $t = \ln(\mu/m_t)$ and $\xi_0$ the boundary condition
$\xi(1/\xi) =\xi_0$. We then define the kick in $\lambda$ for the
models with threshold corrections as
\be
\Delta \lambda = \lambda(t_\star^{\rm SM}) - \lambda_\star^{\rm SM}, \quad
{\rm for} \; \xi_0 = \xi_0^{\rm SM}.
\ee
For our run of 500 simulations, the average kick is upwards
$\langle \Delta \lambda \rangle = 1.5 \times 10^{-3}$ with standard
deviation $\sigma = 3.9 \times 10^{-3}$; the average absolute kick
size is $\langle |\Delta \lambda| \rangle = 2.3 \times 10^{-3}$.  If
the kick is upwards, or downwards but not so large, plateau inflation
is still possible. The value $\xi_0$, which is a free parameter, has
to be adjusted with respect to the reference model, to fit the power
spectrum \eref{xival}. However for large kicks downwards this is no
longer possible, and the potential is too steep for all
$\xi$-values. The critical kick dividing the successful models from
the unsuccessful ones is
\be
\Delta \lambda^{\rm crit} = - 1.1 \times 10^{-3}
\quad
\Rightarrow
\quad
\lambda_\star^{\rm SM} + \Delta \lambda^{\rm crit} = -1.7 \times 10^{-4}
\ee
As mentioned, we do not find any examples of hilltop inflation in our
500 simulations with threshold corrections, in contrast to the pure SM
running.  One can either generate a kick, with respect to the
reference model, by changing the boundary conditions at the
electroweak scale (e.g. changing the EW top/Higgs mass) or by turning
on threshold corrections. Adding thus a kick to the reference model,
the CMB power spectrum constraint is no longer satisfied; we retune
$\xi_0$ to fit the CMB data.  Fig.~\ref{F:kick_xi} shows the result,
it plots $\xi_0$ for downwards kicks $\Delta \lambda <0$.  The
reference model is again $m_t = 171$ GeV, $m_h = 125.5$ GeV, that is
the highlighted line in Table \ref{Table:higgs}.  In this plot the
green line correspond to pure SM running and different values of the
EW top mass, the red line for SM running and different values of the
EW Higgs mass (corresponding to the results in Table
\ref{Table:higgs}), and the blue line for fixed top and Higgs mass but
a kick generated by threshold corrections.  Inflation near the maximum
only happens in the first two cases for the small kick interval where
$\xi_0$ increases again (i.e. where the red and blue line increase).

It matters whether the kick is produced by EW boundary conditions or
by threshold corrections. In the 2nd case, inflation is possible for
larger kick values.  This can be understood as follows.  For SM
running without threshold corrections, changing $\xi_0$ mainly affects
the size of the power spectrum, but it has only a small effect on the
running.  In contrast, for the set-up with threshold corrections,
changing $\xi_0$ will both affect the power spectrum and the running.
Indeed, $\mu \sim 1/\xi_0$ is the scale where the kick is
produced. For a smaller $\xi_0$ this happens at a higher scale, where
the value of $\lambda(\mu)$ is smaller and since the size of the kick
is proportional to $\lambda(\mu)$, this results in a smaller kick.
Hilltop inflation is only possible if the curvature near the maximum
is tuned small.  This depends on the details of the potential.  It is
no surprise that this gives slightly different results for SM running,
and SM including threshold corrections, even for a similarly sized kick.

No matter what the exact form of the threshold corrections is, if the
kick is not too large ($\Delta \lambda + \lambda_*^{\rm SM} >0)$
inflation takes place on the flat plateau with universal predictions
for the observables.  The larger $\lambda_*^{\rm SM}$ is without
threshold corrections, the larger kick is needed to disrupt inflation,
which is only possible for large Wilson coefficients of the
non-renormalizable operators. Consider for example the first line in
Table \ref{Table:higgs}, with
$\lambda_*^{\rm SM} = 6.3 \times 10^{-3}$. Also for this case we did
500 simulations with random Wilson coefficients, choosing
$c^{\rm max} = 20$ with
$c_i = {\rm random}[-c^{\rm max},c^{\rm max}]$. We found that 61 out
of 500 times the downwards kick was large, and inflation no longer
possible; we found no examples of hilltop inflation.

Choosing natural values for the Wilson coefficients
$c^{\rm max} \sim 1$ the effect of the threshold corrections on the
running is small.  Nevertheless, there might be additional sources of
threshold corrections. If they only affect the potential via
modifications of the running, our results apply: (except from some
possible fine-tuned cases near a maximum) inflation takes place on the
plateau and the observables have universal values.  Our choice of
higher order terms in the Lagrangian \eref{L_threshold} with
$c^{\rm max} \sim10$ can be viewed as a (specific) parameterization of
the kick in $\lambda$ due to all possible threshold corrections.

Finally we would like to comment on the possibility of Higgs inflation
near an inflection point, as has been discussed in the literature
\cite{critical1,critical2}. Close to the Planck scale
% the U(1) coupling increases sharply as it nears the Landau pole. 
the potential may develop a second minimum. For fine-tuned boundary conditions the
maximum and minimum merge into an inflection point with
$V_h = V_{hh} =0$, where inflation can take place. Such solutions only
exist for relatively small non-minimal coupling $\xi = \O(10)$; this
is because the renormalization scale is bounded $\mu < 1/\sqrt{\xi}$,
and only for small $\xi$ large enough scales can be reached where the
Landau pole becomes important. Because of the large scales involved,
inflection point inflation can give rise to a large gravitational wave
signal (these models were motivated by the BICEP results). In our
numerics we did not search for this possibility, and it is not
included in our results.

%%%%%%%%%%%%%%%%%%%%%%%%%%%%%%%%%%%%%%%%%%%%%%%%%%%%%%
%%%%%%%%%%%%%%%%%%%%%%%%%%%%%%%%%%%%%%%%%%%%%%%%%%%%%%
%%%%%%%%%%%%%%%%%%%%%%%%%%%%%%%%%%%%%%%%%%%%%%%%%%%%%%
\section{Conclusions}
\label{s:conclusions}

In Higgs inflation the unitarity cutoff, signaling the breakdown of
the effective theory, is well below the Planck scale and introducing
an UV completion is demanded by the consistency of the theory. This
raises the question how sensitive the CMB predictions are to the UV
completion above the cutoff scale. In this paper we have shown that as
long as the UV corrections do not affect the inflaton potential at
tree level but only enter at loop level via corrections to the
renormalization group equations, the inflationary predictions are
(almost) unaffected. Indeed, as we proved analytically in section
\ref{s:flat}, to leading order in the slow roll expansion all
dependence on the running cancels, and thus the predictions are
insensitive to threshold corrections. The spectral index and
tensor-to-scalar ratio are exactly the same as for the classical, tree
level potential, which is in excellent agreement with data.

The inflationary predictions are universal if inflation takes place on
the flat plateau of the potential.  However, it may happen that due to
the running of the couplings the potential develops a maximum.
Inflation near the maximum will depend on the details of the RGE
evolution and thus on the UV completion. The perturbative expansion
used in section \ref{s:flat} does not capture this case, and we used a
numerical analysis to also study the possibility of hilltop inflation,
where we parameterized the threshold corrections by a specific set of
higher order operators in the Lagrangian \eref{L_threshold}. Our
numerical analysis confirms our analytical results for inflation on
the flat plateau of the potential. We further found that hilltop
inflation is a possibility, but it only happens for very fine-tuned
boundary conditions (the top/Higgs mass at the electroweak scale, and
the Wilson coefficients of the non-renormalizable operators).  Indeed,
for our run of 500 simulations with randomly chosen Wilson
coefficients (taken large enough, such that the effect on the running
is appreciable --- see section \ref{s:max} for more details), we found
382 times plateau inflation, and 118 times inflation was spoiled as
the potential became unstable at low field values and the
corresponding maximum was too steep to support 60 efolds of inflation.
In this run, we did not find a single instance of hilltop inflation.
We conclude that, apart from the very fine-tuned case of inflation
near the maximum, if inflation happens, the predictions are the same
as those derived from the classical potential \eref{universal}.

A previous study of threshold correction to HI has been done in
\cite{cliffnew}. They concluded that Higgs inflation is extremely
sensitive to the UV completion, which was modelled by the same set of
higher order operators \eref{L_threshold} \cite{trott}. We expect that
the difference is mainly due to the choice of the renormalization
prescription. While \cite{cliffnew} allows for both prescription 1 and
2 in their numerical analysis, we showed analytically (and confirmed
numerically) that for our choice of $\mu$, which has been discussed
extensively in \eref{s:mu}\footnote{As argued in \eref{s:mu} and in
  \eref{A:RGimproved}, prescription 1 is the only consistent
  renormalization scale parametrization to study the RG improved
  potential in the Einstein frame.}, such dependence does not arise in
general.  Furthermore, there are some slight differences in the
numerical implementation, for example the set of RGEs for the
inflationary regime, and the parameterization of the unitarity
cutoff. However, the main conclusion that inflation on the flat
plateau of the potential is insensitive to UV physics, does not depend
on these choices.  As mentioned above, we do find deviations from the
universal predictions if inflation occurs near a maximum of the
potential.  We studied numerically the fine-tuned parameter space for
hilltop inflation, which depends sensitively on the boundary
conditions as well as on the UV completion --- and thus also on the
specifics of the numerical implementation.  For standard model
inflation, without threshold corrections, our numerical results agree
with earlier work \cite{bezrukov3,kyle}.

We conclude with a small remark. It is well known that for the central
values of the electroweak scale top and Higgs mass the Higgs potential
becomes unstable at $\phi \sim 10^{10}$ GeV \cite{disc1, disc2,
  branchina,branchina2, archil, alexss,kniehl}, well before the
potential flattens in Higgs inflation. The top/Higgs mass values
separating a stable from an unstable Standard Model Higgs potential
are close to those separating Higgs inflation from models where
inflation is not possible. There are small differences with respect to
SM running (without a non-minimal coupling), because 1) we include
threshold effects, 2) we run until the inflationary scale and not the
Planck scale, and 3) the RGE equations get modified in the mid and
large field regime.  As expected the measured Higgs and top masses
\cite{Aad:2014eva} are surprisingly close to the border separating the
regions where the Higgs boson can or cannot be the inflaton.

\section*{Acknowledgments}
We thank Sander Mooij, Enrico Morgante, Subodh Patil, Mike Trott and
Cliff Burgess for very helpful discussions.  The authors are funded by the
Netherlands Foundation for Fundamental Research of Matter (FOM) and
the Netherlands Organisation for Scientific Research (NWO).

%%%%%%%%%%%%%%%%%%%%%%%%%%%%%%%%%%%%%%%%%%%%%%%%%%%%%%
%%%%%%%%%%%%%%%%%%%%%%%%%%%%%%%%%%%%%%%%%%%%%%%%%%%%%%
%%%%%%%%%%%%%%%%%%%%%%%%%%%%%%%%%%%%%%%%%%%%%%%%%%%%%%
\appendix
\section{Effective action and renormalization group improvement}
\label{A:RGimproved}

In Higgs inflation the canonically normalized field $h$ and the field
$\phi$ appearing in the Einstein frame action \eref{LE} are related by
a non-trivial field space metric \eref{canonical}. Although it is
possible to write $h(\phi)$ in closed form
\cite{linde_higgs,GarciaBellido:2008ab}, this relation can only be
inverted in certain limits. Therefore, the potential in terms of the
canonical field $h$ can not be expressed in an analytical form over
the whole field domain, and it is often more convenient to work with
the $\phi$-field (as we did, for example, in section
\ref{s:inflation}).

The RG improved potential is usually defined in terms of the canonical
field.  In this appendix we will show that the usual procedure of
substituting each coupling with its running counterpart can also be
used for the non-canonical $\phi$ field, and the RG improved potential
is obtained as\footnote{That this is possible is not immediately
  obvious. In the large field regime
  $\phi = \frac1{\sqrt{\xi}}\e^{h/\sqrt{6}}$; as this relation depends
  explicitly on a coupling, the potential in terms of the canonical
  field has a different coupling dependence (it only depends on the
  combination $\lambda/\xi^2$), and at first sight it might seem that
  the prescription of making all couplings running \eref{V_RGI}
  differs when done in terms of $h$ or $\phi$. We also did the
  calculation of the inflationary predictions in terms of the
  canonical field, and at leading order in the $\delta$-expansion
  found identical results to those presented in section
  \ref{s:inflation}.}
 %%\textcolor{red}{(once we look at them in the regime where the potential can be written analytically in terms of $h$, i.e. large $\xi$ limit and $\phi\gg1/\xi$.)}}
\begin{equation}
V_{E}=\frac{\lambda\phi^{4}}{\left(1+\xi\phi^{2}\right)^{2}}
\,\,\longrightarrow\,\,\frac{\lambda(t)\phi^{4}}{\left(1+\xi(t)\phi^{2}\right)^{2}}.
\label{V_RGI}
\end{equation}
We first quickly review how the RG improved potential can be defined
for canonical fields, and then generalize to the case with non-trivial
field space metric.

\subsection{Canonical kinetic sector}

Consider first the SM without the non-minimal coupling, the Higgs
kinetic term is canonical.

The effective action $\Gamma[\phi_{\rm cl}]=W[J]-\int J\phi_{\rm cl}$, with
$\phi_{\rm cl}=\delta W/\delta J$
describes the quantum corrected dynamics of the background field,
since
\begin{equation}
\frac{\delta\Gamma[\phi_{\rm cl}]}{\delta\phi_{\rm
    cl}}\bigg|_{\bar{\phi}_{\rm cl}}=0,
\qquad\bar{\phi}_{\rm cl}=\langle\Omega|\phi|\Omega\rangle\equiv\langle\phi\rangle,
\end{equation}
i.e. the vacuum expectation value is given by minimizing the effective
action. $\Gamma[\phi_{\rm cl}]$
has the following form
\begin{equation}
\Gamma[\phi_{\rm cl}]=S_{r}[\phi_{\rm cl}]+\Delta S_{c}[\phi_{\rm cl}]\,+\Gamma^{\rm 1-loop}+\Gamma^{\rm 2-loop}+...
\end{equation}
The first term is the classical renormalized action, $\Delta S_{c}$
contains the counterterms and the third term represent the one loop
correction%
\footnote{We only consider one-loop corrections in the inflationary regime.}, 
\begin{equation}
\Gamma^{\rm 1-loop}=\frac{i}{2}\int d^{4}x\,\sum_{i}(-1)^{F_{i}}S_{i}{\rm Tr} \ln(D_{i}+m_{i}^{2}(\phi_{\rm cl}))\label{eq:gamma 1-loop11}
\end{equation}
$S_{i}$ counts the degrees of freedom of each particle with mass
$m_{i}$, $F_{i}$ is $1$ for fermions and $0$ for bosons. $\Gamma$ is
finite (physical amplitudes are derived from it), as the 
infinities from the loop contributions are eliminated by the counterterms.

Usually one is interested  in finding the space-time
independent vacuum state. Thus $\phi_{\rm cl}$ is taken constant, 
and the one loop contribution to $\Gamma$ can easily be computed
since the operators inside the log become diagonal in momentum representation.
%\begin{equation}
%\Gamma^{\rm 1-loop}\supset-3i\rm Trln(i\partial+m_{t}^{2}(\phi_{\rm cl}))+\frac{i}{2}\rm Trln(\square+m_{h}^{2}(\phi_{\rm cl}))+..\label{eq:gamma one loop}
%\end{equation}
% For example if $m_{h}(\phi_{\rm cl})$ is constant over space-time 
%\begin{equation}
%i\rm Trln(\square+m_{h}^{2})=i\int d^{d}x\int\frac{d^{d}k}{(2\pi)^{4}}ln(-k^{2}+m_{h}^{2})\,\longrightarrow\,\int\frac{d^{4}x}{64\pi^{2}}m_{h}^{4}\left(\ln\frac{m_{h}^{2}}{\mu^{2}}-\frac{3}{2}\right)\label{eq:simple computation}
%\end{equation}
%where the standard $\texttt{\ensuremath{\bar{MS}}}$ renormalization
%scheme has been understood%
%\footnote{i.e. the counter terms are used only to remove the part proportional
%to $\bar{\epsilon}^{-1}=(\epsilon^{-1}-\gamma+\ln4\pi)$, with $\epsilon=(4-d)/2$.%
%} and $\mu$ is the renormalization scale. 
The effective action for a constant background field reduces then to the effective potential given by
the tree level contribution plus the well known Coleman-Weinberg corrections \cite{CW}
\begin{align}
\Gamma[\phi_{\rm cl}]&=-\int d^{4}x\, V_{\rm eff}[\phi_{\rm cl}]
\nn\\
&=-\int d^{4}x\left[\, V_{\rm tree}(\phi_{\rm cl})+\frac{1}{64\pi^{2}}\sum_{i}(-1)^{F_{i}}S_{i}m_{i}^{4}(\phi_{\rm cl})\[\ln \(\frac{m_i^2(\phi)}{\mu^2}\) -c_i \]\right]+..
\end{align}
in the $\overline{\rm MS}$ renormalization scheme. Here $V_{\rm tree}$ is
the potential at tree level, $\mu$ is the normalization scale, and
$c_i = 3/2$ for fermions and scalars and $c_i =5/6$ for gauge bosons.
Minimizing the effective potential gives the vev for a constant background
scalar field.

The perturbative expansion breaks down when the logs in the CW
potential become large.  This problem can be avoided by rewriting the
effective action as a formal solution of the RG equation. In fact, due
to the invariance of the theory with respect to the renormalization
procedure $\Gamma$ satisfies the Callan-Symanzik equation
\cite{Callan:1970yg,Symanzik:1970rt}
%This problem can be avoided by using the RG
%improved potential
%%which resums all the logs. 
%The improved action is
%a formal solution of the  Callan-Symanzik equation, which encodes that
%the action cannot depend on the arbitrary normalization scale $\mu$:
\begin{equation}
\left(\mu\frac{\partial}{\partial\mu}+\beta_{i}\frac{\partial}{\partial g_{i}}-\gamma\int d^{4}x\,\phi_{\rm cl}\frac{\delta}{\delta\phi_{\rm cl}}\right)\Gamma[\phi_{\rm cl},g_{i},\mu]=0.
\end{equation}
The formal solution is given by straightforwardly applying the method of characteristics \cite{Ford:1992mv}
\begin{equation}\label{improved}
V_{\rm eff}(\phi_{\rm cl},g_{i},\mu)=V_{\rm eff}(\phi(t),g_{i}(t),\mu(t))\equiv V(t),
\end{equation}
 with
\begin{equation}
\phi(t)=\rho(t)\phi_{\rm cl},\quad
\frac{dg_{i}(t)}{dt}=\beta_{i}(g_{j}(t)),\quad
\mu(t)=\mu e^{t}, \quad
\frac{d\ln\rho(t)}{dt}=-\gamma(g_{j}(t)).
\label{eq:1}
\end{equation}
Here $g_{i}$ represent the generic couplings and $\gamma$ is the
anomalous dimension of the Higgs field. We also assume the initial
conditions $\rho(0)=1,\,\,\,g_{i}(t)=g_{i},\,\,\,\mu(0)=\mu$. 
%Using the running coupling is equivalent to reorganize perturbation theory in such a way that the dominant logs are resummed; this can be made explicit by expanding $g_i(\phi_{\rm cl})$ in $(\phi_{\rm cl}/\mu)$ \cite{scrucca}. The resummation works best, and large logarithms are avoided, if $\mu$ is chosen such that the log in the CW potential is minimized, i.e. $\mu^2 \sim m_i^2(\phi_{\rm cl})$

The power of the RG is the fact that we can choose the functional form
of $t=t(\phi)$ in such a way that the perturbation series for $V(t)$
converges more rapidly than the one for $V(0)$. This can be made
explicit by choosing $\mu(t)$ such that the logs in the CW potential
are minimized, i.e. $\mu(t)^2 \sim m_i^2(\phi_{\rm cl})$.

%MP: i don't understand this:
%The usefulness ofSince, by construction, $V(t)$ is independent on $t$ we can choose
%the functional form of the function $\mu(t)$, i.e. we choose $t$ such that
%the perturbation series for $V(t)$ converge more rapidly minimizing
%the logarithm expansion in the potential. 

During inflation the background field is rolling down its potential,
i.e. $\phi_{\rm cl}(t)$ is not constant. As a result the masses
appearing in expressions like (\ref{eq:gamma 1-loop11}) are not
constant and the 1-loop contributions must be calculated in
spacetime-dependent perturbation theory
\cite{Fraser:1984zb,Iliopoulos:1974ur,mp}. The effective action
assumes the generic form
\begin{equation}
\Gamma[\phi_{\rm cl}]=-\int d^{4}x\,\[\frac{1}{2}Z(\phi_{\rm cl})\partial_{\mu}\phi_{\rm cl}\partial^{\mu}\phi_{\rm cl}+V_{\rm eff}[\phi_{\rm cl}]+...\]
\end{equation}
where the dots are for higher derivatives terms $\sim Y(\phi_{\rm cl})(\partial\phi_{\rm cl})^{4}+..$
that we can safely neglect in the slow roll approximation, and $Z=Z(\phi_{\rm cl},g_{i},\mu)$.
%\footnote{One way among the others to understand this form for the effective
%action it is setting $\phi_{\rm cl}=\bar{\phi}_{cl}+\tilde{\phi}(x)$,
%with $\bar{\phi}_{cl}$ constant in (\ref{eq:gamma 1-loop11}). Expanding
%the operators inside the log's in powers and derivatives of $\tilde{\phi}$
%and recollecting the coefficient in front we obtain expressions for
%$Z(\phi_{\rm cl})$ as well as for the other higher derivatives terms
%(ref. calculation of higher derivative terms in the one loop effective
%lagrangian and functional methods and perturbations theory). Another
%way it is to compute that explicitly for $\phi_{\rm cl}=\phi_{\rm cl}(t)$
%as it has been done in (Marieke paper).%}.
Applying the Callan-Symanzik
equation to the kinetic term gives, after some integrations by part,
the following expression
\begin{equation}
\left(\mu\frac{\partial}{\partial\mu}+\beta\frac{\partial}{\partial\lambda}-\gamma\left(2+\phi_{\rm
      cl}\frac{\partial}{\partial\phi_{\rm
        cl}}\right)\right)Z(\phi_{\rm cl},g_{i},\mu)=0.
\label{eq:callam for kinetic}
\end{equation}
The formal solution can be written as 
\begin{equation}
Z(\phi_{\rm
  cl},g_{i},\mu)=Z\left(\phi(t),g_{i}(t),\mu(t)\right)\rho^{2}(t)\equiv
Z_{\rm eff}(t).
\label{eq:solution kinetic}
\end{equation}
where $\{\phi(t),\rho(t),g_{i}(t),\mu(t)\}$ are given in \eref{eq:1},
with the same initial conditions. In the leading order approximation%
\footnote{Since also $Z$ depends on a series of logarithms, the best choice
of $t$ to minimize them in the effective potential is to set $Z$
to one at the leading order.%
} 
\begin{equation}
Z_{\rm eff}(t)\approx\rho^{2}(t)=e^{-2\int_{0}^{t}\gamma(t')dt'}
\label{Zeff}
\end{equation}
and the improved effective action becomes
\begin{equation}
\Gamma=-\frac{1}{2}\int \[Z_{\rm eff}(t)(\partial_{\mu}\phi_{\rm cl})^{2}+V(t)\].
\end{equation}
As usual, it is convenient to use a canonical field redefinition,
i.e. $Z_{\rm eff}^{1/2}d\phi_{\rm cl}=d\phi_{\rm can}$. This is useful
for two reasons: the equations become simpler and the gauge dependence
in the potential is significantly reduced (see the discussion in
Ref. \cite{Espinosa:2015qea}). Consider for example the tree level potential
$V_{\rm tree}=\lambda\phi_{\rm cl}^{4}/4$.  The improved effective action
is
\begin{equation}
\Gamma=-\int\[\frac{Z_{\rm eff}(t)}{2}(\partial_{\mu}\phi_{\rm cl})^{2}+\frac{\lambda(t)}{4}\rho^{4}(t)\phi_{\rm cl}^{4}+..\]=-\int\[\frac{1}{2}(\partial_{\mu}\phi_{\rm can})^{2}+\frac{\lambda(t)}{4}\phi_{\rm can}^{4}+..\]
\end{equation}
where in the last step we used \eref{Zeff}.  This explains why, with
proper choice of the function $t$ (or equivalently $\mu(t)$), the
effect of the renormalization group can be summarized, from an
operative point of view, in taking the action and making the couplings
running.  In this example
$\lambda(t)=d\beta_{\lambda}(\lambda(t))/dt$.

\subsection{Non canonical kinetic sector}

Let us now see how the previous discussion can be generalized to
the non-minimal kinetic terms in Higgs inflation. The Einstein frame
effective action is
\begin{equation}
\Gamma[\phi_{\rm cl}]=-\int d^{4}x\,\[ \frac{1}{2}Z(\phi_{\rm
  cl})\gamma_{\phi\phi}(\phi_{\rm cl},\xi)\partial_{\mu}\phi_{\rm
  cl}\partial^{\mu}\phi_{\rm cl}+V_{\rm eff}[\phi_{\rm cl}]+{\rm h.o.} \]
\end{equation}
The fermion, gauge boson, Higgs and Goldstone boson masses are given
in \eref{mass}, and the field space metric $\gamma_{\phi\phi}$ in
\eref{canonical}.  Even though the kinetic terms are non-canonical,
one can define the improved effective action as a
formal solution of the RG equations. The result now is that
$Z'\equiv Z(\phi_{\rm cl})\gamma_{\phi\phi}(\phi_{\rm cl},\xi)$
satisfies an equation of the form (\ref{eq:callam for kinetic}) and
its solution can be rewritten, like in (\ref{eq:solution kinetic}), as
\begin{equation}
Z'_{\rm
  eff}(t)=Z\left(\phi(t),g_{i}(t),\mu(t)\right)\gamma_{\phi\phi}\left(\phi(t),\xi(t)\right)\rho^{2}(t)\,
%\overset{\rm LO}{\approx}
\approx
\,\gamma_{\phi\phi}(t)\rho^{2}(t).
\end{equation}
The improved effective action takes the form 
\begin{equation}
\Gamma=-\int
d^{4}x\,\[\tfrac{1}{2}\rho^{2}(t)\gamma_{\phi\phi}(\phi(t),\xi(t))(\partial_{\mu}\phi_{\rm
  cl})^{2}
+V_{\rm eff}(\phi(t),g_{i}(t),\mu(t)\],
\end{equation}
with $g_{i}$ labeling all SM model couplings plus the non-minimal
coupling $\xi$. $V(t)$ is the effective potential for Higgs inflation rewritten as a solution of the RG equation, i.e \eref{improved}.   Now proceed exactly as before in order to rewrite the
improved action in terms of a canonical field. Let us do that in two
steps. First use the following field redefinition
\begin{equation}\label{first redefinition}
d\tilde{\phi}_{\rm can}=\rho(t)d\phi_{\rm cl}.
\end{equation}
Then for $\phi(t)$ we obtain 
\begin{equation}
\phi(t)=\rho(t)\phi_{\rm cl}=\exp\left(-\int_{0}^{t}\gamma(t)dt\right)\rho^{-1}(t)\tilde{\phi}_{\rm can}\approx\tilde{\phi}_{\rm can},
\end{equation}
and the effective lagrangian becomes ($\Gamma\equiv \int d^{4}x \L_{\rm eff}$)
\begin{equation}\label{improved2}
\L_{\rm eff}=-\frac{1}{2}\gamma_{\phi\phi}\left(\tilde{\phi}_{\rm can},g_{i}(t),\mu(t)\right)(\partial_{\mu}\tilde{\phi}_{\rm can})^{2}-V_{\rm eff}\left(\tilde{\phi}_{\rm can},g_{i}(t),\mu(t)\right).
\end{equation}
Then define the canonical field via
\begin{equation}\label{second redefinition}
d\phi_{\rm can}=\sqrt{\gamma_{\phi\phi}}d\tilde{\phi}_{\rm
  can}.
\end{equation}
To connect to the notation of the rest of this paper set
$\tilde{\phi}_{\rm can}\equiv\phi$ and $\phi_{\rm can}=h$. 
% MP: this story absent in previous subsection, comes out of nowhere
The improved potential in \eref{improved2} is the Higgs inflation quantum improved potential (eq. \eref{improved}) after the field redefinition \eref{first redefinition}, i.e.
\begin{equation}
V_{\rm eff}=\frac{\lambda(t)\phi^{4}}{(1+\xi(t)\phi)^{2}}+\frac{1}{64\pi^{2}}\sum_{i}S_{i}m_{i}^{4}(\phi,g_{i}(t))\left[ \ln \left(\frac{m_{i}^{2}(\phi,g_{i}(t))}{\mu(t)}\right)-c_{\rm i}\right].
\end{equation}
The optimal choice for the renormalization scale, which kills the (dominant) logs of the top and gauge boson
loop contributions in the CW potential, is given by
\begin{equation}
\mu(t)=m_{t}(\mu\sim EW)\e^t \sim\frac{\phi}{\Omega(t)}=\frac{\phi}{\sqrt{1+\xi(t)\phi^{2}}}
\end{equation}
where $m_{t}(\mu\sim EW)$ is the top mass measured at the Electroweak
scale. This is \eref{mu2}, which is our choice for the renormalization scale.
It follows that the RG improved potential becomes 
\begin{equation}
V_{\rm eff}(\phi)\simeq\frac{\lambda(t(\phi))\phi^{4}}{(1+\xi(t(\phi))\phi^{2})^{2}},
\end{equation}
where $\phi=\phi(h)$ through \eref{second redefinition}.
%With this choice our effective potential has been improved as 
%\begin{equation}
%V_{\rm eff}(\phi)=\frac{\lambda(t(\phi))\phi^{4}}{(1+\xi(t(\phi))\phi^{2})^{2}}
%\end{equation}
% with $t=\ln(\phi/m_{t}\Omega)$ and we can safely neglect the log
%contribution.

\section{CMB parameters at higher order in $\delta$}
\label{A:perturb}

In this appendix we compute the perturbation spectrum at second order
in the slow roll expansion. At this order the results do depend on the
running. We check that there is no accidental cancellations or terms
blowing up, and that the leading order results are indeed the dominant terms.

In order to compute the CMB parameters $(n_{s},r)$ at second order
in $\delta=1/(\xi\phi^{2})$  we need the slow roll parameter  $\eta$
at 2nd order, and 
$\epsilon$ at 3rd order (as $\sqrt{\eps}$ enters the integral for the
number of efolds). Define 
\begin{equation}
\mathcal{K}\equiv\frac{V_{h}}{V}=\frac{1}{h_{\phi}}\frac{V_{\phi}}{V}=\sqrt{\frac{8}{3}}\frac{\left(1+\frac{\beta_{\lambda}}{4\lambda}\right)\delta(\delta+1)}{(\frac{\delta+1}{6\xi}+1)^{\frac{1}{2}}\left(\delta+1+\tfrac{\beta_{\xi}}{2\xi}\right)}
\end{equation}
Then the slow roll parameters can be written as
\begin{align}
\epsilon&=\frac{1}{2}\mathcal{K}^{2}=\frac{8}{3}\left(1+\frac{\beta_{\lambda}}{4\lambda}\right)\frac{\delta^{2}(\delta+1)^{2}}{A\,
          B^{2}},\label{eq:epsilonfull} \\
\eta&=\frac{V_{hh}}{V}=\frac{1}{V}\frac{d}{dh}\left(\mathcal{K}V\right)=\mathcal{K}_{h}+\mathcal{K}^{2}=\frac{1}{h_{\phi}}\mathcal{K}_{\phi}+2\epsilon,\label{eq:eta second order}
\end{align}
where we have defined 
\begin{equation}
A=\(\frac{\delta+1}{6\xi}+1\),\qquad B=\(1+\frac{\beta_{\xi}}{2\xi}+\delta\).
\end{equation}
Now $\mathcal{K}_{h}=h_{\phi}^{-1}\mathcal{K}_{\phi}$
with $h_{\phi}=\sqrt{\gamma_{\phi\phi}}$ given in terms of the field
space metric \eref{canonical}. Explicitly
\begin{equation}
\mathcal{K}_{h}=\frac{1}{h_{\phi}}\left(\frac{\partial\mathcal{K}}{\partial\delta}\delta_{\phi}+\frac{\partial\mathcal{K}}{\partial f_{j}}\frac{df_{j}}{dt}\frac{dt}{d\phi}\right),\label{eq:kchi}
\end{equation}
where $f_{j}\equiv\{\lambda,\beta_{\lambda},\xi,\beta_{\xi}\}$ and
thus $df_{j}/dt\equiv\{\beta_{\lambda},\beta'_{\lambda},\beta_{\xi},\beta'_{\xi}\}$.
The  full non-expanded slow roll parameters are  given by  \eref{eq:epsilonfull} and\footnote{For completeness $\frac{1}{3}\left(1+\frac{\beta_{\lambda}}{4\lambda}\right)\frac{\delta^{2}(\delta+1)^{2}}{A\, B^{3}}\left[\frac{\beta_{\xi}}{\xi^{2}}\left(\frac{B}{6A}+\beta_{\xi}\right)-\frac{\beta'_{\xi}}{\xi}\right]\equiv O\left(\frac{\beta_{\xi}}{\xi^{2}},\frac{\beta'_{\xi}}{\xi}\right)$.%
} 
\begin{equation}\begin{split}\label{eq: etafull}
\eta=\,&\frac{V_{hh}}{V}=\frac{4}{3}\left(1+\frac{\beta_{\lambda}}{4\lambda}\right)\left(\frac{\delta^{2}(\delta+1)^{2}}{A\, B^{2}}\left(1+\frac{1}{12\xi}\frac{B}{A}+2\left(1+\frac{\beta_{\lambda}}{4\lambda}\right)\right)-\frac{\delta(\delta+1)(2\delta+1)}{A\, B^{2}}\right)\\[1mm]
&+\frac{2}{3}\left(\frac{\beta'_{\lambda}}{4\lambda}-\frac{\beta_{\lambda}^{2}}{4\lambda^{2}}\right)\frac{\delta^{2}(\delta+1)^{2}}{A\, B^{2}}+O\left(\frac{\beta_{\xi}}{\xi^{2}},\frac{\beta'_{\xi}}{\xi}\right)
\end{split}\end{equation}
For $\lambda_{\max}=-{\beta_{\lambda}}/{4}$ or equivalently $F=0$,
$\epsilon$ reduces to zero (extremum of the potential) while $\eta$ reduces to \eref{sr_crit}.

In order to compute the number of efolds we expand $\epsilon$ at
third order in $\delta$, which means we need to expand $\mathcal{K}$
at second order in $\delta$, 
\begin{equation}
\mathcal{K}\approx\, k_{1}\delta+k_{2}\delta^{2}+O(\delta^{3})\label{eq:dude}.
\end{equation}
Then $\epsilon$ is given by $\epsilon=\epsilon_{0}\delta^{2}+\epsilon_{1}\delta^{3}+O(\delta^{4})$
with $\epsilon_{0}=k_{1}^{2}/2$ and $\epsilon_{1}=k_{1}k_{2}$. $N_{\star}$
becomes 
\begin{equation}
N_{\star}=\int_{\phi_{{\rm end}}}^{\phi_{\star}}\frac{1}{\sqrt{2(\epsilon_{0}\delta^{2}+\epsilon_{1}\delta^{3}+..)}}h_{\phi}\, d\phi,
\end{equation}
with
\be\epsilon\approx\frac{1}{2}k_{1}^{2}\delta_{{\rm
    end}}^{2}=1\quad\implies\quad\phi_{{\rm
    end}}\approx\left(\frac{4}{3}(1+\tfrac{1}{6\xi})\right)^{1/4}\left(\frac{F_{{\rm
        end}}}{\xi_{{\rm end}}}\right)^{1/2}\approx\left(\frac{F_{{\rm
        end}}}{\xi_{{\rm end}}}\right)^{1/2}.  
\ee
To
understand which terms are important we expand the integrand,
i.e. $N_{\star}\equiv\int f$, schematically as
$f\sim O(1/\sqrt{\delta})+O(\sqrt{\delta})+O(\delta)+...$, where
$\int O(1/\sqrt{\delta})\propto\phi_{\star}^{2}\,\,;\int
O(\sqrt{\delta})\propto \ln(\phi_{\star});\,\,\int
O(\delta)\propto1/\xi\phi_{\star}$.
The results can be written in term of $\delta_{\star}$ as\footnote{
  Following the arguments below \eref{eq:exp} we neglect the
  implicit $\phi$ dependence of the couplings and $\beta$-functions.}
\begin{equation}
N_{\star}\approx a_{1}\frac{1}{\delta_{\star}}+a_{2}\ln\delta_{\star}+\mathcal{C},\label{eq:riferimento}
\end{equation}
 with
\begin{equation}
a_{1}=\frac{3}{4F_{\star}},\quad
a_{2}=\frac{3}{4F_{\star}}\left(\frac{1}{1+\frac{1}{6\xi_{\star}}}+\frac{\beta_{\xi\star}}{2\xi_{\star}(1+\frac{\beta_{\xi\star}}{2\xi_{\star}})}\right),\quad
\mathcal{C}=-\frac{3}{4F_{\star}}\xi_{\star}\phi_{{\rm end}}^{2}+a_{2}{\rm ln}(\xi_{\star}\phi_{{\rm end}}^{2}).
\end{equation}
Now rewrite (\ref{eq:riferimento}) as 
\begin{equation}
\delta_{\star}=\frac{a_{1}}{N_{\star}}+\frac{a_{2}\delta_{\star}{\rm ln}\delta_{\star}}{N_{\star}}+\frac{\mathcal{C}\delta_{\star}}{N_{\star}},\label{eq:ite}
\end{equation}
 which can be solved iteratively. At leading order $\delta_{\star}=a_{1}N_{\star}^{-1}+O(N_{\star}^{-2})$
and plugging that back in (\ref{eq:ite}) gives
\begin{equation}
\delta_{\star}=\frac{a_{1}}{N_{\star}}+\frac{a_{2}a_{1}}{N_{\star}^{2}}{\rm ln}\left(\frac{a_{1}}{N_{\star}}\right)+\frac{\mathcal{C}a_{1}}{N_{\star}^{2}}+O(N_{\star}^{-3}).\label{eq:deltastar}
\end{equation}
Then $\epsilon$ evaluated at horizon exit becomes at second order
\begin{equation}
\epsilon_{\star}\approx\frac{4}{3}\left(1+\frac{1}{6\xi_{\star}}\right)F_{\star}^{2}\frac{a_{1}^{2}}{N_{\star}^{2}}=\frac{3}{4}\frac{1}{N_{\star}^{2}}\left(1+\frac{1}{6\xi_{\star}}\right)
\end{equation}
Expanding \eref{eq: etafull} at second order in $\delta$ and using
\eref{eq:deltastar} (we set also $\xi\gg1$ for simplicity and we
neglect $O(\beta_{\xi}/\xi^{2},\beta'_{\xi}/\xi)$) we obtain  
\begin{equation}
\eta_{\star}\approx-\frac{1}{N_{\star}}+\frac{3}{2N_{\star}^{2}}-\frac{3}{4N_{\star}^{2}}\frac{1}{F_{\star}}\left(1-\frac{1}{2F_{\star}}\left(\frac{\beta'_{\lambda\star}}{4\lambda_{\star}}-\frac{\beta_{\lambda\star}^{2}}{4\lambda_{\star}^{2}}\right)-{\rm ln}\left(\frac{\xi_{{\rm end}}}{\xi_{\star}}\frac{F_{\star}}{F_{{\rm end}}}N_{\star}\right)-\frac{\xi_{\star}}{\xi_{{\rm end}}}F_{{\rm end}}\right)+O(N_{\star}^{-3}).
\end{equation}
 Therefore the CMB parameters are given by the expressions 
\begin{align}
n_{s}&=1+2\eta_{\star}-6\epsilon_{\star}\nn\\
&\approx1-\frac{2}{N_{\star}}-\frac{3}{2N_{\star}^{2}}-\frac{3}{2N_{\star}^{2}F_{\star}}\left(1-\frac{1}{2F_{\star}}\left(\frac{\beta'_{\lambda\star}}{4\lambda_{\star}}-\frac{\beta_{\lambda\star}^{2}}{4\lambda_{\star}^{2}}\right)-{\rm
  ln}\left(\frac{\xi_{{\rm end}}}{\xi_{\star}}\frac{F_{\star}}{F_{{\rm
  end}}}N_{\star}\right)-\frac{\xi_{\star}}{\xi_{{\rm end}}}F_{{\rm
  end}}\right), \nn\\
r&=16\epsilon_{\star}\approx\frac{12}{N_{\star}^{2}}.
\end{align}
Turning off the running $F_{\star}=1,\,\beta_{i}=0$, the tree level
result are recovered at second order in $N_{\star}^{-1}$,
i.e. $n_{s}=1-\frac{2}{N_{\star}}-\frac{3}{N_{\star}^{2}}+...$.  The
spectral index $n_{s}$ feels the effect of the running only at second
order.  This dependence goes as $F_{\star}^{-1}$. Note, however, that
for values of $F_{\star}$ close to zero the $\delta$ expansion breaks
down, and we can no longer trust our analytical results. This is
exactly the case where the potential has a maximum and we study the
problem numerically.

%%%%%%%%%%%%%%%%%%%

%%%%%%%%%%%%%%%%%%%%%%%%%%%%%%%%%%%%%%%%%%%%%%%%%%%%
%%%%%%%%%%%%%%%%%%%%%%%%%%%%%%%%%%%%%%%%%%%%%%%%%%%%
%%%%%%%%%%%%%%%%%%%%%%%%%%%%%%%%%%%%%%%%%%%%%%%%%%%%

%\end{document}
\bibliographystyle{utphys}
\bibliography{biblio}

\end{document}